\documentclass[acmsmall,screen]{acmart}

\AtBeginDocument{%
  \providecommand\BibTeX{{%
    Bib\TeX}}}

\setcopyright{acmlicensed}
\copyrightyear{2025}
\acmYear{2025}
\acmDOI{XXXXXXX.XXXXXXX}

\usepackage[ruled,vlined,linesnumbered]{algorithm2e}
\usepackage{amsthm}
\usepackage{amsmath}
\usepackage{amsfonts}
\usepackage{amssymb}
\usepackage{mathtools}
\usepackage{subcaption}
\usepackage{soul}
\usepackage[table,xcdraw]{xcolor}
\usepackage{multirow}
\usepackage{multicol}
\usepackage[utf8]{inputenc}
\usepackage{graphics}
\usepackage{wrapfig}
\usepackage{svg}
\usepackage{tcolorbox}
\usepackage{balance}
\usepackage{enumitem}
\usepackage{comment}
\usepackage{textcomp}
\usepackage{xspace}
\usepackage{upgreek}
\usepackage{tikz}
\usetikzlibrary{arrows.meta, positioning, calc, fit}

\def\BibTeX{{\rm B\kern-.05em{\sc i\kern-.025em b}\kern-.08em
    T\kern-.1667em\lower.7ex\hbox{E}\kern-.125emX}}

\definecolor{darkgreen}{rgb}{0,0.52,0}

\newcommand*\colourcheck[1]{%
  \expandafter\newcommand\csname #1check\endcsname{\textcolor{#1}{\ding{52}}\xspace}%
}
\newcommand*\colourcross[1]{%
  \expandafter\newcommand\csname #1cross\endcsname{\textcolor{#1}{\ding{56}}\xspace}%
}

\newboolean{showcomments}
\setboolean{showcomments}{true}
\ifthenelse{\boolean{showcomments}}
{
	\definecolor{myyellow}{RGB}{255, 228, 26}
	\definecolor{myblue}{RGB}{50, 50, 220}
	\newcommand{\nb}[2]{
		{\sf
			\fcolorbox{myyellow}{yellow}{\scriptsize\textbf{#1}}%
			$\blacktriangleright$%
			{\color{myblue}\fontsize{7pt}{8pt}\selectfont\textbf{#2}}%
		}%
	}
}
{
	\newcommand{\nb}[2]{}
}

\newcommand{\tool}{\textsc{Mimicry}\xspace} %
\newcommand{\deepjanus}{DeepJanus\xspace} %
\newcommand{\head}[1]{\noindent\textbf{#1.}}

\renewcommand*{\equationautorefname}{Equation}
\def\equationautorefname~#1\null{(#1)\null}

\RestyleAlgo{ruled}
\SetKwComment{Comment}{\# }{}
\newlength\mylen

\DeclareFontFamily{OT1}{pzc}{}
\DeclareFontShape{OT1}{pzc}{m}{it}{<-> s * [1.10] pzcmi7t}{}
\DeclareMathAlphabet{\mathpzc}{OT1}{pzc}{m}{it}

\begin{document}

\title{Targeted Deep Learning System Boundary Testing}

\author{Oliver Wei{\ss}l}
\orcid{0009-0008-7575-0187}
\email{weissl@fortiss.org}
\affiliation{%
  \institution{Technical University of Munich}
  \city{Garching near Munich}
  \country{Germany}
}
\affiliation{%
  \institution{fortiss}
  \city{Munich}
  \country{Germany}
}

\author{Amr Abdellatif}
\email{amr.abdellatif@tum.de}
\orcid{0009-0004-0225-2428}
\affiliation{%
  \institution{Technical University of Munich}
  \city{Garching near Munich}
  \country{Germany}
}

\author{Xingcheng Chen}
\email{xingcheng.chen@tum.de}
\affiliation{%
  \institution{Technical University of Munich}
  \city{Garching near Munich}
  \country{Germany}
}
\email{xchen@fortiss.org}
\orcid{0009-0002-0861-4093}
\affiliation{%
  \institution{fortiss}
  \city{Munich}
  \country{Germany}
}
\author{Giorgi Merabishvili}
\orcid{0009-0000-0314-7487}
\email{gmerabi@ncsu.edu}
\affiliation{%
  \institution{North Carolina State University}
  \city{Raleigh}
  \country{USA}
}  

\author{Vincenzo Riccio}
\orcid{0000-0002-6229-8231}
\email{vincenzo.riccio@uniud.it}
\affiliation{%
  \institution{University of Udine}
  \city{Udine}
  \country{Italy}
}

\author{Severin Kacianka}
\orcid{0000-0002-2546-3031}
\email{kacianka@fortiss.org}
\affiliation{%
  \institution{fortiss}
  \city{Munich}
  \country{Germany}
}

\author{Andrea Stocco}
\orcid{0000-0001-8956-3894}
\email{andrea.stocco@tum.de}
\affiliation{%
  \institution{Technical University of Munich}
  \city{Garching near Munich}
  \country{Germany}
}
\email{stocco@fortiss.org}
\affiliation{%
  \institution{fortiss}
  \city{Munich}
  \country{Germany}
}

\renewcommand{\shortauthors}{Wei{\ss}l et al.}

\begin{abstract}
Evaluating the behavioral boundaries of deep learning (DL) systems is crucial for understanding their reliability across diverse, unseen inputs. Existing solutions fall short as they rely on untargeted random, model- or latent-based perturbations, due to difficulties in generating controlled input variations. In this work, we introduce \tool, a novel black-box test generator for fine-grained, targeted exploration of DL system boundaries. \tool performs boundary testing by leveraging the probabilistic nature of DL outputs to identify promising directions for exploration. It uses style-based GANs to disentangle input representations into content and style components, enabling controlled feature mixing to approximate the decision boundary. We evaluated \tool's effectiveness in generating boundary inputs for five widely used DL image classification systems of increasing complexity, comparing it to two baseline approaches. Our results show that \tool consistently identifies inputs closer to the decision boundary. It generates semantically meaningful boundary test cases that reveal new functional (mis)behaviors, while the baselines produce mainly corrupted or invalid inputs. Thanks to its enhanced control over latent space manipulations, \tool remains effective as dataset complexity increases, maintaining competitive diversity and higher validity rates, confirmed by human assessors.
\end{abstract}


\received{08 May 2025}
\received[revised]{19 September 2025}
\received[accepted]{28 September 2025}

\maketitle

\section{Introduction}\label{sec:introduction}

The increasing dependence on Deep Learning (DL) systems for both everyday tasks and critical sectors~\cite{schutze2008introduction} makes rigorous testing for these systems a relevant topic~\cite{riccio2020testing,zhang2020machine}.
The concept of fault in DL systems is more complex than in traditional software~\cite{riccio2020testing}. 
Even if the code that builds the DL network is bug-free, the trained DL model may still deviate from the expected behavior due to faults introduced during the training phase, such as the misconfiguration of learning parameters or the use of an unbalanced or non-representative training set~\cite{2020-Humbatova-ICSE}. 
In data-intensive software systems, such as DL systems, faults often stem from the large, high-dimensional input space, which requires the generation of test data that accurately captures the complexity and diversity of the validity domain, i.e., the portion of the input space for which the system is expected to operate~\cite{riccio2020testing}.

Test generation techniques have been developed to induce misbehaviors in DL systems~\cite{riccio2020testing,zhang2020machine,riccio2020deepjanus,deeproad,deeptest,pei2017deepxplore,sun2018testing}. 
However, the objectives of these techniques are often quite different. Some techniques focus on finding adversarial examples~\cite{guo2018dlfuzz, zhang2020deepsearch, kurakin2018adversarial, croce2020minimally}, while other solutions aim to achieve high failure exposure and/or high values of DL-specific adequacy metrics, such as neuron~\cite{ma2018deepgauge} or surprise coverage~\cite{kim2019guiding}, or explore the decision boundaries of the DL system~\cite{riccio2020deepjanus,kang2020sinvad}.

In particular, DL boundary testing targets regions of the input space where small input variations can lead to misbehaviors. Boundary inputs are crucial for evaluating the DL system's reliability, as they often expose how it handles edge cases, transitions between operational domains, and critical decision-making regions. 
In traditional software systems, boundary testing is typically targeted. For example, consider a Java method \texttt{sum(x, y)} that adds two integers, where each parameter ranges from $-2^{32}$ to $2^{31}$. A boundary testing strategy for this method would include inputs such as the minimum allowed value ($-2^{32}$), its immediate successor ($-2^{32} + 1$), an arbitrary in-range value (e.g., 100), the maximum allowed value ($2^{31}$), and its immediate predecessor ($2^{31} - 1$). 
Since there are two parameters, this targeted approach yields only $5^2 = 25$ combinations, covering edge behaviors that are most likely to reveal bugs.
In contrast, boundary testing for DL systems is challenging due to high-dimensional, unconstrained input spaces (e.g., images) and unclear input space partitions. As such, existing solutions such as \deepjanus~\cite{riccio2020deepjanus} and Sinvad~\cite{sinvad-tosem} rely on untargeted boundary testing strategies. These are commonly driven by evolutionary algorithms that generate diverse inputs without any explicit consideration of specific source or target classes. While these methods can uncover unexpected behaviors, they tend to be inefficient and unfocused, as they treat the entire input space uniformly rather than concentrating on regions near critical decision boundaries.
However, DL models inherently learn decision boundaries between classes. For instance, in a digit classification task, given an image of class 5, the DL model may assign high probabilities to both the classes 5 and 6, reflecting the probabilistic nature of the model's output rather than a definitive classification~\cite{10.1007/s10664-023-10393-w}. This suggests the model is uncertain between these two classes, making inputs from class 6 promising candidates for generating boundary cases. Thus, it is potentially more effective to focus testing on inputs near the classifier's decision boundary between classes that share some features like 5 and 6, rather than sampling randomly across unrelated classes. Despite this potential, targeted boundary testing in DL systems remains largely unexplored.

While researchers have explored various approaches, existing solutions have key limitations that hinder their effectiveness in boundary testing of complex DL systems. An example is \deepjanus~\cite{riccio2020deepjanus}, an input generation technique that relies on an abstract representation of the input domain (i.e., a model) to generate test cases. However, such domain models are typically unavailable for complex, feature-rich datasets such as ImageNet. 
Although recent advances in generative AI have addressed the lack of explicit input models, current techniques for generating inputs in the latent space of DL models~\cite{kang2020sinvad,dola2024cit4dnn,DBLP:journals/corr/abs-2001-11055,dunn2021exposing} either do not target boundary inputs, or they offer limited control over the generation process due to the use of a single, entangled latent vector perturbed by random noise~\cite{sinvad-tosem}, thereby severely constraining the ability to navigate the latent space. 

In this paper, we propose a technique to explore the boundary of DL systems in the latent space of style-based generative adversarial networks.
The key idea involves leveraging a style transfer architecture that automatically learns the separation of high-level features (e.g., shape) from lower-level ones (e.g., texture). While this architecture is primarily used for the generation of new, highly diverse datasets of complex inputs, in this work, we leverage the scale-specific control on the synthesis of disentangled latent factors for boundary testing of DL systems.

Our technique, implemented in a tool called \tool \cite{replication-package}, uses style-specific interpolation operations to find boundary inputs. 
\tool uses a conditional StyleGAN~\cite{karras2019style} model trained to learn the class-wise visual characteristics of a given image dataset across all its inputs. 
StyleGAN maps latent vector inputs to an intermediate latent vector, which controls the image style at various granularity levels in the generative process. 
The main idea of \tool involves the systematic mutation of the pre-defined set of latent vectors between source and target inputs using scale-specific 
interpolation and assessing the impact of these modifications in the image space. 
Moreover, \tool facilitates the targeted generation of boundary inputs by leveraging model confidence scores. Given a source input, \tool identifies the boundary target as the class with the second-highest predicted confidence from the DL system. It then establishes a closed feedback loop between the DL model under test and the StyleGAN network to guide input synthesis. Specifically, \tool employs StyleGAN to generate representative samples of the target class and manipulates the latent representation of the source input to incorporate features of these target samples, thereby adjusting its visual characteristics toward the decision boundary.

We have evaluated the effectiveness of \tool on five popular image classification datasets with increasing complexity (MNIST~\cite{mnist}, FashionMNIST~\cite{fashionmnist}, SVHN~\cite{svhn}, CIFAR-10~\cite{cifar}, ImageNet~\cite{deng2009imagenet}) to assess its robustness across a diverse range of visual patterns and challenges, using self and pre-trained WideResNet~\cite{zagoruyko2016wide} as DL systems under test. 
Additionally, we compare the effectiveness of \tool against the model-based \deepjanus~\cite{riccio2020deepjanus} and the generative-based Sinvad~\cite{sinvad-tosem}. Our experiments demonstrate that \tool consistently identifies inputs close to the decision boundary while maintaining a high validity rate and label preservation rate, as evaluated by human assessors. 
Moreover, \tool surpasses both \deepjanus  and Sinvad in both quantitaive and qualitative metrics, especially when increasing data complexity.
Our paper makes the following contributions:

\begin{description}[noitemsep]
\item [Technique.] To the best of our knowledge, \tool is the first targeted boundary testing technique for DL systems. Our approach is implemented in the publicly available tool \tool~\cite{replication-package} and is based on a disentangled latent space representation that ensures high controllability.
\item [Evaluation.] An empirical study shows that \tool is more effective than existing model-based and generative-based techniques in various quality metrics, including higher effectiveness, validity, and label-preservation rates. 
\end{description}
\section{Background}\label{sec:background}

\subsection{Testing Objectives for DL Systems}

\begin{figure}[t]
    \centering
    \begin{subfigure}{0.3\textwidth}
        \centering
        \includegraphics[width=\linewidth]{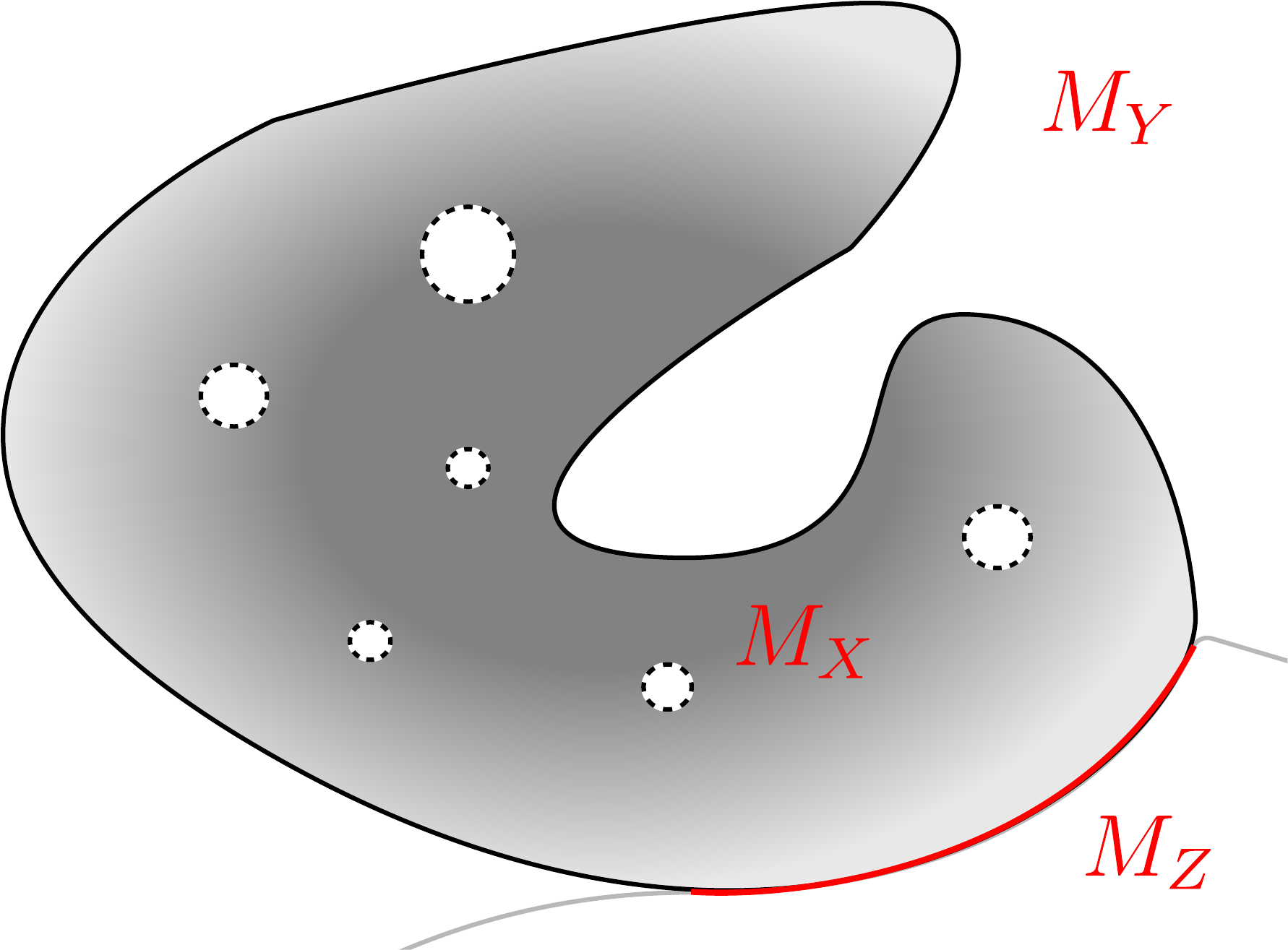}
        \caption{\scriptsize{Targeted Boundary Testing.}}\label{fig:manif_c}
    \end{subfigure}
    \begin{subfigure}{0.3\textwidth}
        \centering
        \includegraphics[width=\linewidth]{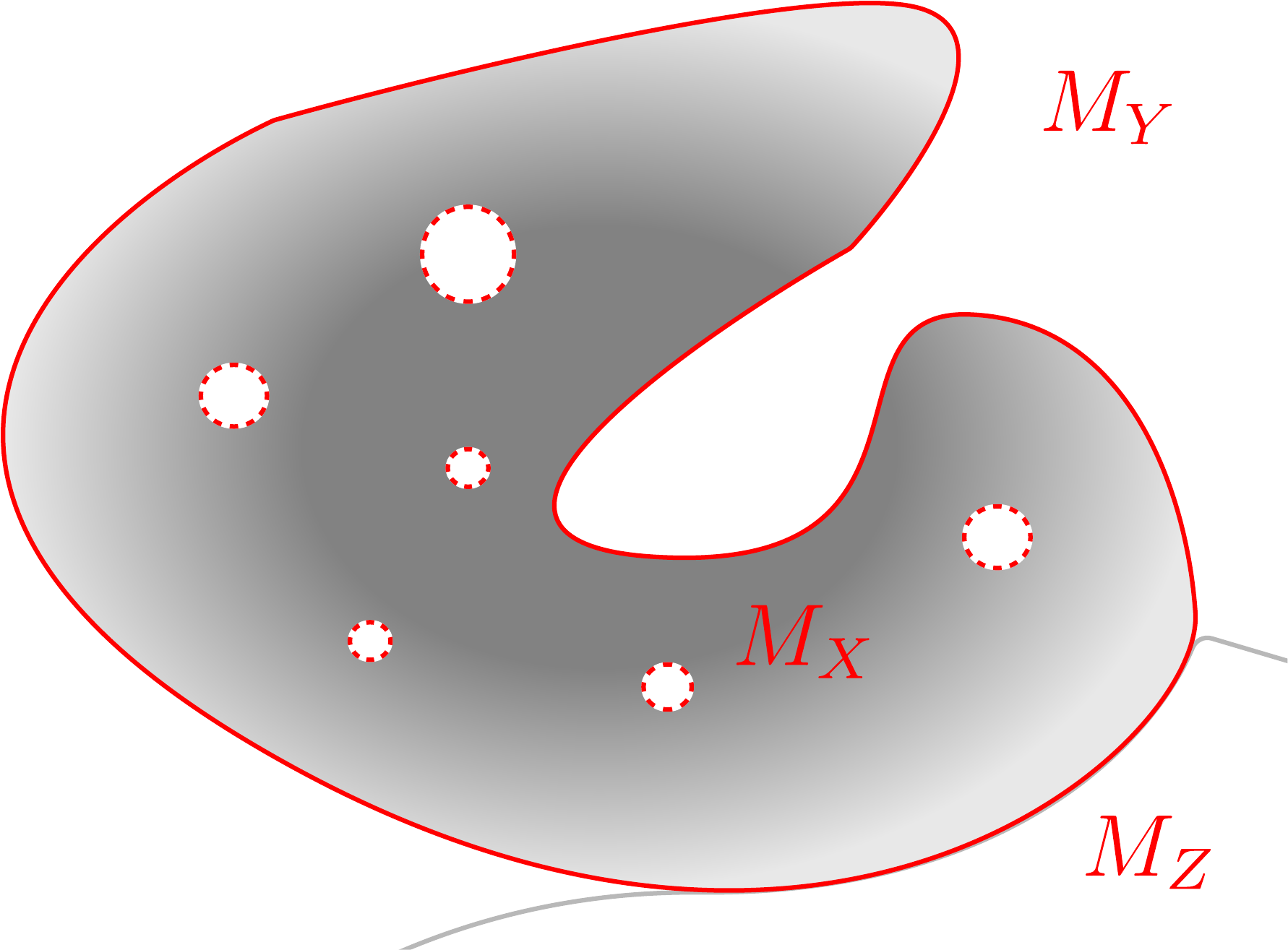}
        \caption{\scriptsize{Untargeted Boundary Testing.}}\label{fig:manif_d}
    \end{subfigure}
     \caption{Different types of boundary testing objectives for DL systems.}
    \label{fig:manif}
\end{figure}

Testing methodologies to highlight behavior in DL models can have vastly different objectives. The distinctions are often unclear in the related literature. Therefore, we define key terms and specify the experimental domain. We illustrate these differences using a classifier manifold $M$, which encodes all possible classifier decisions, mapping from a high dimensional input space to the lower dimensionality space $M$.
Within this manifold, distinct regions $M_\bullet$ (sub-manifolds) exist, corresponding to each classifiable class, respectively. In \autoref{fig:manif} the main sub-manifold for class $X$, $M_X$ is in focus, showing its ``boundaries'' to other regions on $M$ and internal boundaries which symbolize adversarial regions~\cite{7727230}.
 
In this work, we explore boundary testing, a subset of functional testing, which targets the \textit{generalizability} aspect of the SUT by generating functionally new inputs. Some approaches such as DeepXplore~\cite{pei2017deepxplore}, DLFuzz~\cite{guo2018dlfuzz}, and DeepTest~\cite{deeptest} involve raw input manipulation techniques that modify/corrupt the original inputs (e.g., pixels). These techniques do not generate new functional inputs as they produce changes in the original inputs and are therefore suitable to test the deficiencies in robustness of the DL system~\cite{2023-Riccio-ICSE,2025-Maryam-ICST}, such as the discovery of adversarial regions. In contrast, the generation of functional tests focuses on creating new inputs that deviate from the original training distribution. These inputs target the long-tail problem of DL testing~\cite{zhang2024systematicreviewlongtailedlearning}, testing the DNN's ability to generalize to novel, unseen scenarios. Instances of functional test generators are the model-based approaches like \deepjanus~\cite{riccio2020deepjanus}, DeepHyperion~\cite{zohdinasab2021deephyperion} and DeepMetis~\cite{2021-Riccio-ASE} or latent space manipulation techniques like SINVAD~\cite{kang2020sinvad,sinvad-tosem}, CIT4DNN~\cite{dola2024cit4dnn}, and RBT4DNN~\cite{mozumder2025rbt4dnnrequirementsbasedtestingneural}.

\subsection{Boundary Testing for DL Systems}

Boundary testing identifies input samples near decision boundaries, where the classifier assigns equal, or near-equal, probabilities to multiple classes~\cite{riccio2020deepjanus, kang2020sinvad}. 
The decision boundary of a classifer can be inferred from the predicted logits, where the theoretical boundary would be a perfect equilibrium in confidences between $n$ classes ($n>1$). In addition, boundary testing can be either targeted or untargeted.
The goal of targeted boundary testing is to converge to the boundary between the origin class and a specified target class (e.g., $M_Z$ in \autoref{fig:manif_c}), while the goal of the untargeted case assumes no predetermined target class (\autoref{fig:manif_d}).

While existing techniques such as DeepJanus~\cite{riccio2020deepjanus} and Sinvad~\cite{kang2020sinvad} focus on untargeted boundary testing, in our work, we focus on targeted boundary testing, with the goal of automatically retrieving inputs that are ambiguous in prediction, without restrictions on the input differences.

\subsection{Style-Based Generative Adversarial Networks}
Generative Adversarial Networks (GANs) are DL models designed to learn the statistical distribution of a training dataset, allowing the synthesis of new samples that are representative of the learned distribution~\cite{goodfellow2020generative}. 
GANs involve jointly training a pair of networks that compete with each other. This approach is based on game theory and is implemented using two neural networks. A first neural network, called the generator, aims to produce realistic images, while a second neural network, called the discriminator, acts as an expert that receives both fake and real (authentic) images and aims to distinguish between them. In this way, the generator improves its ability to produce realistic images to fool the discriminator, which can be leveraged for test generation~\cite{dola2023input26, dunn2021exposing}.

StyleGAN~\cite{karras2019style, karras2020analyzing, karras2021alias, sauer2022styleganXL} extends the GAN architecture to introduce new methods for controlling the image synthesis process. Unlike traditional GANs, StyleGAN enables style control at multiple levels within the network. The proposed changes to the generator model involve the use of a mapping network to map points in the initial latent space to an intermediate latent space.  This intermediate latent space controls the strength of the image features at various scales in the generator model, inspired by the style transfer literature~\cite{Huang_2017_ICCV}. This architectural change, combined with the noise injected directly into the network, enables the automatic, unsupervised separation of high-level attributes from stochastic variations in the generated images, which we exploit for boundary testing.

\section{Methodology}\label{sec:methodology}

\begin{figure}
    \centering
    \includegraphics[width=0.7\linewidth]{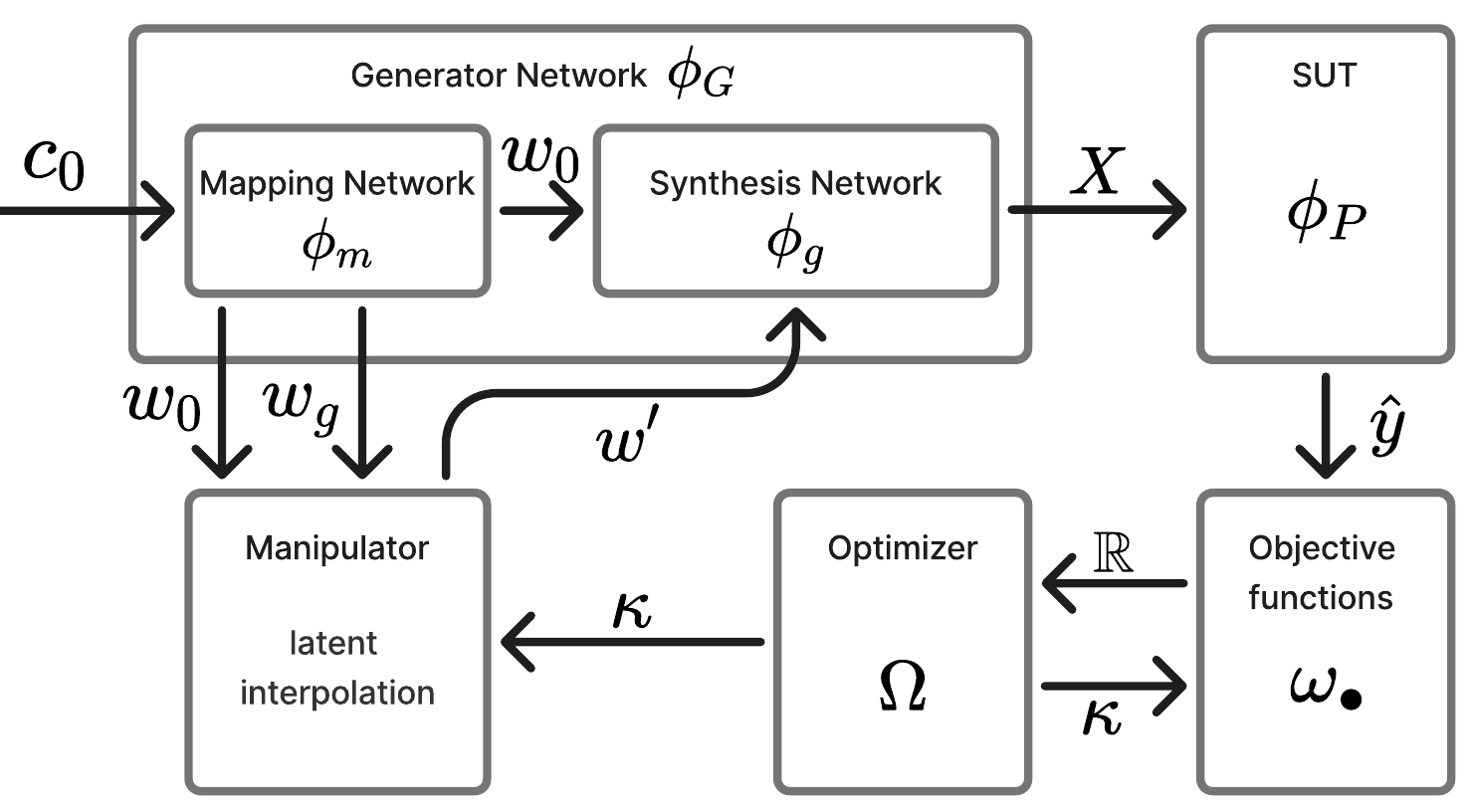}
    \caption{\tool component interactions./}
    \label{fig:cap}
\end{figure}

\tool is a black-box approach\footnote{A recent survey classifies methods that need access to both the training and test datasets of a learned component as \textit{data-box} methods. However, to avoid confusion, we refer to these as black-box methods, since they do not utilize any internal information from the model itself~\cite{riccio2020testing}.} that leverages StyleGANs to generate boundary inputs through \textit{targeted} optimization. Concretely, \tool uses four separate components (\autoref{fig:cap}):

\begin{itemize}
    \item \textbf{SUT}: The system under test, for which the behavioral aspects should be explored.
    \item \textbf{Manipulator}: A component that adapts inputs to the SUT by generating new data points based on a given strategy.
    \item \textbf{Optimizer}: Responsible for providing strategies to the manipulator by evaluating the quality or fitness of previous strategies.
    \item \textbf{Objectives}: Metrics that quantify the quality of solutions, guiding the optimizer's search.
\end{itemize}

\tool operates by identifying boundary inputs based on feedback from the SUT's (e.g., a DL classifier) predictions. These boundary inputs are generated via latent space manipulations in a StyleGAN model trained on the same data as the classifier. These manipulations are driven by strategies optimized according to 
a set of objective functions (\autoref{fig:cap}).

The process is initialized by specifying the number of optimization generations and the initial class $c_0$ to test. An initial latent vector $w_0$ is sampled from the StyleGAN and used to generate the corresponding image $X$. If the predicted class of this image does not match the intended initial class, a new sample is drawn, as the current input is already failure-inducing. Once a valid initial image is obtained, its latent vector (seed) is iteratively optimized toward a boundary candidate by applying linear interpolations with another (target) latent vector.

\tool's targeted nature lies in its treatment of boundary discovery: instead of focusing solely on maximizing misclassification~\cite{riccio2020deepjanus, kang2020sinvad, sinvad-tosem}, it identifies the second most probable class as the target. This initial bias toward a specific class 
allows \tool to exploit the proximity to a specific decision boundary. The target seed $w_g$, together with the original latent vector, is then manipulated according to a strategy $\kappa$, which is optimized by the optimizer using defined objectives $\omega_{\bullet}$. In the following sections we will describe each component of \tool in more detail.

\subsection{System under Test}

The first component used by \tool is the SUT. In this work we target DL classification systems, as they inherently involve the notion of classes---and consequently, boundaries between classes, which is a requirement for performing boundary testing.

We denote the classifier as $\phi_P$, a trainable map from image input to the set of possible classes $C$. Specifically, the output is a vector of class probabilites $\mathbb{R}^i \xrightarrow{\phi_P}\mathbb{R}^{||C||}$.

Here, the superscript $i$ denotes the shape of the input $i$. We can think of the classification operation as positioning our input on the classifier manifold, with the location being the predicted class confidences. On this classifier manifold, \tool aims to finding boundaries between regions of different classes. The boundaries are regions where the classifier's confidence $\hat{y} \in \mathbb{R}^{||C||}$ is equidistant between two or more classes in the set $C_t$ \autoref{eq:eqd}. Note here that $\sum \hat{y} = 1$. 

\begin{equation}
     \forall c \in C_t, \quad \hat{y}_c = \frac{1}{||C_t||}.\label{eq:eqd}
\end{equation}

\subsection{Manipulator}

To find boundary cases, \tool manipulates latent vectors from a conditional StyleGAN model. StyleGANs were specifically chosen because their disentangled latent space offers greater control over manipulations. Unlike traditional GAN architectures, where a single (noise) latent vector is used to generate outputs ~\cite{goodfellow2020generative}, StyleGAN uses a latent vector that passes through an additional network called the mapping network $\phi_m$, which consists of multiple fully connected layers. This mapping network ``disentangles'' the latent space by distributing learnt image characterisitcs across the layers of an intermediate latent vector $w$. The number of layers in $w$ depends on the specific StyleGAN architecture, with more layers enabling finer control over image manipulations. In contrast, traditional GANs and VAEs can be seen as having only a single such layer in $w$, which limits the degree of control.

The StyleGAN model generates new images from class information \autoref{eq:grmap} and can be denoted as a composition of a mapping network $\phi_m$ and a synthesis network $\phi_g$ \autoref{eq:gcomp}. 
We denote it as $\phi_m(\cdot)$, where the only explicitly given input is the class information, as $z$ is sampled noise.
\vspace{-2em}
\begin{multicols}{2}
\begin{equation}
    \mathcal{C} \xrightarrow[]{\phi_G} \mathbb{R}^i.  \label{eq:grmap}
\end{equation}\vfill
\begin{equation}
    \phi_G = \phi_g \circ \phi_m.
    \label{eq:gcomp}
\end{equation} 
\end{multicols}

\begin{figure}[tb]
    \centering
    \includegraphics[width=0.9\linewidth]{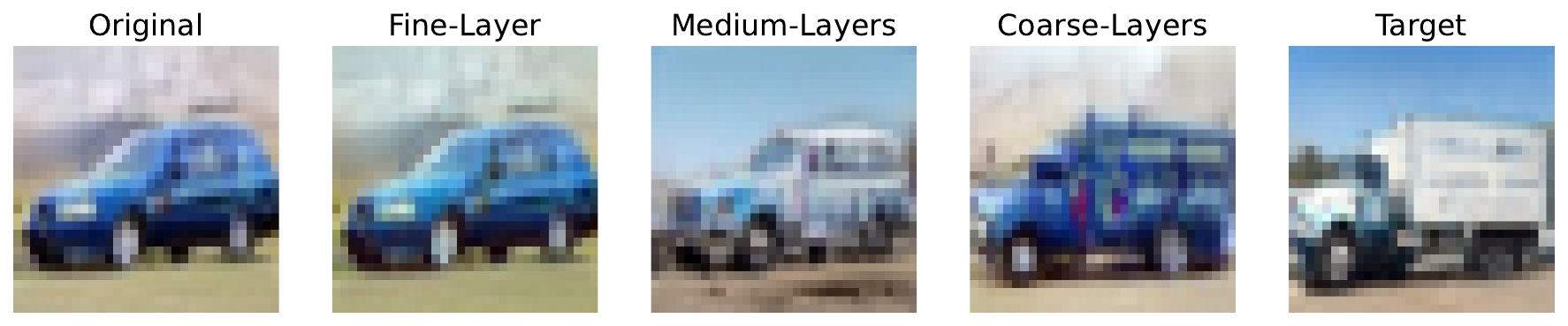}
    \caption{Mixing features of an original image (blue car) with those of a target image (white truck) produces different outputs depending on the latent layers.}
    \label{fig:sgex}
\end{figure}

To get from a class to a generated image in StyleGAN we sample a latent vector $z \sim \mathcal{N}$ which then is processed by the mapping network in combination with the class information to generate an intermediate latent vector $w$ \autoref{eq:gmapd}
The latent vector $w$ is then used for manipulation, as it can be separated into multiple independent layers, depending on the StyleGANs architecture. The advantage of this conversion is that the manipulation of latent vectors $z$ may produce erratic changes in the image, as observed in previous studies~\cite{kang2020sinvad,2023-Riccio-ICSE,dunn2021exposing,dola2024cit4dnn}. 

\begin{equation}\label{eq:gmapd}
    \mathcal{C} \times z \xrightarrow{\phi_m} w \xrightarrow{\phi_g} \mathbb{R}^i.
\end{equation}

Additionally, as noted by Karras et al. \cite{karras2020analyzing}, the rank of manipulations in the intermediate latent space of StyleGAN2 allows for control at different levels of granularity in the generated images. Specifically, the first three layers of the intermediate latent vector $w$ tend to control coarse features such as overall shape and perspective. The next four layers influence medium-scale attributes, including textures and finer structural details. Finally, the last layers typically affect only color schemes, making them responsible for the most fine-grained manipulations. An example of these effects can be seen in \autoref{fig:sgex}, where layers of the original latent vector are mixed with elements of a differing target vector (here car vs. truck in CIFAR-10).
While the exact layer assignments are specific to StyleGAN2, this behavior is consistent across all StyleGAN architectures, as they share the same type of intermediate latent space $w$, albeit with varying numbers of layers \cite{sauer2022styleganXL, karras2019style, karras2020analyzing, karras2021alias}.
The intermediate latent vectors $w$ can are subsequently synthesized into an image using the synthesis network $\phi_g$.

\begin{figure}[tb]
    \centering
    \begin{subfigure}[b]{0.49\linewidth}
        \centering
        \includegraphics[width=0.8\linewidth]{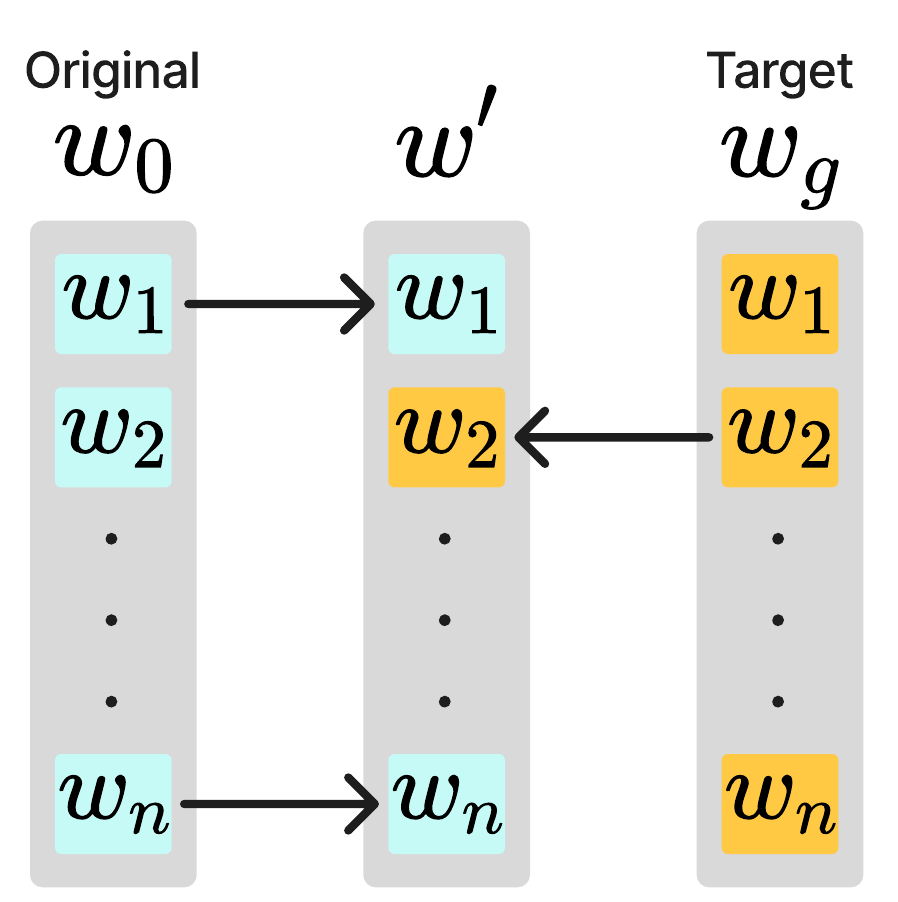}
        \caption{Style Mixing.}
        \label{fig:stylemix}
    \end{subfigure}
    \hfill
    \begin{subfigure}[b]{0.49\linewidth}
        \centering
        \includegraphics[width=0.8\linewidth]{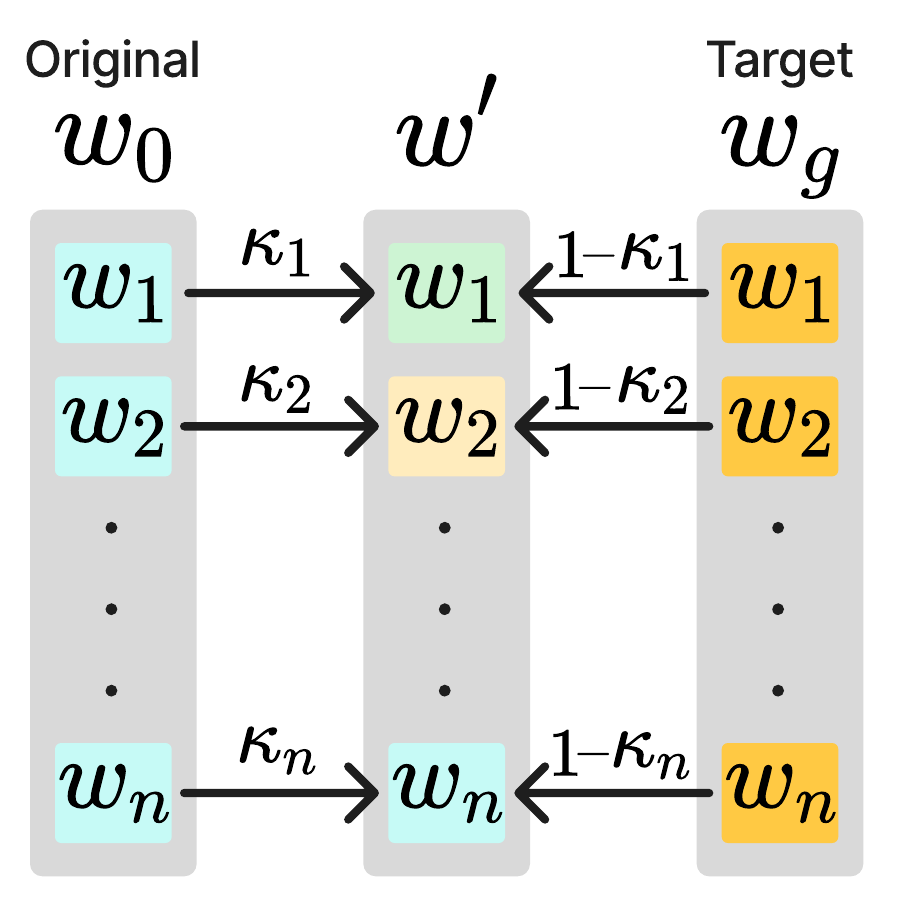}
        \caption{Style Interpolation.}
        \label{fig:latentinterp}
    \end{subfigure}
    \caption{Latent manipulation approaches.}
    \label{fig:methods}
\end{figure}

\subsection{Optimizer}

We combine the latent vector of the initial class $w_0$ and the latent vector of a target class $w_g$ into a new latent vector $w' = \kappa w_0 + (1-\kappa) w_g$ (see \autoref{fig:latentinterp}), where $\kappa$ is the manipulation strategy found by the optimizer. To find adequate manipulation strategies, \tool uses the AGE-MOEA-2 optimizer~\cite{panichella2022improved}, known for its outstanding performance with one or multiple objectives. Contrary to the original style mixing approach of Karras et al.~\cite{karras2019style}, the produced strategies $\kappa$ do not swap individual layers (\autoref{fig:stylemix}), but they are weights for linear interpolation between two layers of the same rank in two latent vectors (\autoref{fig:latentinterp}). This increases the controllability of feature mixing between the source and target seed, which is useful for precise DL boundary assessment.

The latent vectors for interpolation (seeds) in our case are selected as follows: The first seed is of the original class $w_0$, whereas the second seed $w_g$ is dependent on the second most likely prediction of the primary seed by the SUT. This reuse of SUT behavior allows for a more targeted boundary search, as \tool incorporates knowledge of the decision space that is traversed. At the start of optimization, the manipulation strategies $\kappa$ are initialized at random to cover a wide range of manipulations. 
In \autoref{fig:cap}, the Optimizer is denoted as $\Omega$, with its inputs being directly linked to the collection of objective functions used, described in \autoref{sec:objf}.

\subsection{Objectives for Boundary Testing} \label{sec:objf}

For boundary testing, we are interested in regions of the classifier manifold (decision space) where the predicted class probabilities are in equilibrium. When a manipulated input results in such an equilibrium, it indicates that the classifier is unable to decisively distinguish between two or more classes, revealing the presence of a decision boundary.

During classifier training, data points are mapped onto the classifier's manifold with the goal of separating instances from different classes while drawing instances of the same class closer together. This iterative process results in the formation of clusters within the decision space, with the boundaries between these clusters representing the decision boundaries.

\begin{figure}[t]
    \centering
    \includegraphics[width=0.5\linewidth]{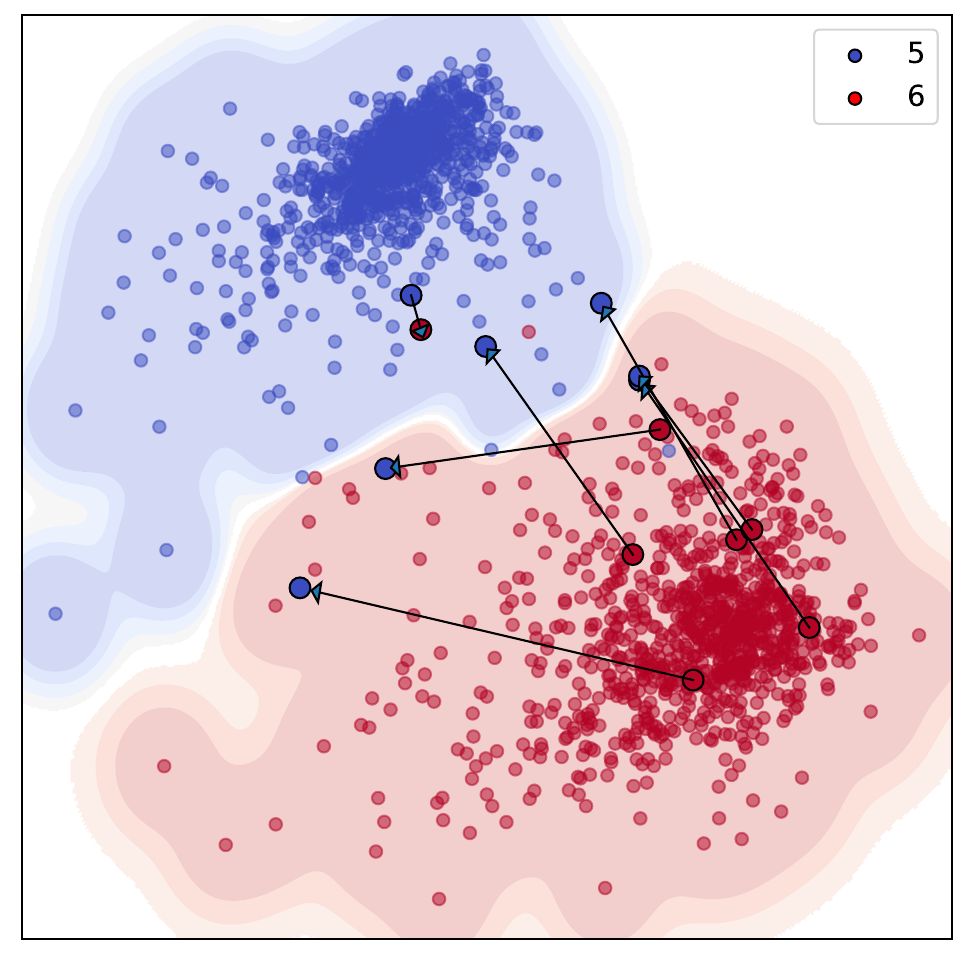}
    \caption{Boundary discovery between classes ``5'' and ``6'' in the MNIST benchmark.}
    \label{fig:bound_manif}
\end{figure}

As discussed in the previous section, \tool traverses this space by manipulating two or more latent vector seeds, allowing us to generate input that are approaching the decision boundary. \autoref{fig:bound_manif} shows different candidate boundary inputs generated by \tool for the classes ``5'' and ``6'' of the MNIST dataset. The small points are images from the original MNIST test set, whereas the larger points are the boundary images generated by \tool, with arrows connecting the source and target seeds. Since \tool uses the second most likely class in the SUTs predictions we have a (dynamic) target of boundary, allowing for more efficient search as additional knowledge of the decision surface is encoded into the procedure. In targeted boundary testing, the targets can either be rigid or dynamic, where the rigid approach would entail that once a target class is established it cannot change.
\tool uses the dynamic approach in targeted boundary testing, allowing for target change through optimization if the classifiers behavior suggest a boundary to a different class might be closer. Allowing the target to change during optimization enables more flexible and efficient boundary search. It helps navigate intersections of multiple class boundaries and escape local minima by shifting focus to closer or more reachable decision regions.

In our experiments, we use two objectives to optimize toward boundary candidates. The first objective function consists in a \textit{dynamic confidence balance}, depicted in \autoref{eq:dcb}. Here $\hat{y}'$ are the predicted class probabilities of the manipulated input $X'$. The subscript $1st$ denotes the index of the primary seed class and the set $J = \mathcal{C}_g$ are all the other elements that should be considered (in our case, we only consider the second most-likely class from the SUT). This function essentially quantifies how similar multiple confidences are to each other and how much weight they have combined against the rest of the confidence values. This means that the more similar the targeted confidence values are and the more confidence they encapsulate in the prediction, the higher the dynamic confidence balance.

\begin{equation}
    \omega_{dcb} : \frac{\sum_{j\in J} |\hat{y}'_1-\hat{y}'_j|}{||J||(\hat{y}'_1 + \sum_{j\in J} \hat{y}'_j)}.\label{eq:dcb}
\end{equation}

In addition to \autoref{eq:dcb}, we also integrate a quality criterion, similar to DeepJanus~\cite{riccio2020deepjanus}. This quality criterion measures the normalized Euclidean distance $d_2$ between strategies (genomes) of an archive $\kappa_A$ and a population $\kappa_P$, to enforce a greater novelty of solutions in genomes. Here we want to maximize sparsity of new genomes, based on the parent population which functions as the archive, therefore we minimize \autoref{eq:gd2}. The search for novelty is commonly used to limit the exploitation of local minima and have a better traversal of the optimizer search space. This novelty measure does not concern the actual generated images, rather their manipulation weights.

\begin{equation}
    \omega_{d2} : 1-min\Big\{\frac{d_2(a,p)}{\sqrt{||a||}}\;\Big|\; (a,p) \in \kappa_A \times \kappa_P\Big\}.\label{eq:gd2}
\end{equation}

\section{Empirical Study}\label{sec:study}

\subsection{Research Questions}
To evaluate the proposed tool, we consider the following research questions:
\\
\textbf{RQ\textsubscript{1} (effectiveness):} How effective is \tool in finding boundary inputs? 
\\
\textbf{RQ\textsubscript{2} (efficiency):} How efficient is \tool in finding boundary inputs?
\\
\textbf{RQ\textsubscript{3} (quality):} To what extent are the inputs generated by \tool valid and label-preserving?
\\
\textbf{RQ\textsubscript{4} (latent space usage):} How is the disentangled latent space used to generate boundary inputs?

RQ\textsubscript{1} assesses whether \tool is able to find test cases close to the boundary, and whether it is able to cover a wide range of different boundary cases with regards to the boundary target. 
RQ\textsubscript{2} evaluates the efficiency in terms of runtime to investigate the potential cost of utilizing \tool. 
RQ\textsubscript{3} studies the quality of the inputs produced by \tool, in terms of validity, as assessed by human evaluators.
RQ\textsubscript{4} involves an internal evaluation to determine how \tool's usage of the disentangled latent space affects the inputs manipulations.

\subsection{Objects of Study} 

\subsubsection{Datasets} 
In our study, we used five image classification datasets, namely MNIST~\cite{mnist}, FashionMNIST~\cite{fashionmnist}, SVHN~\cite{svhn}, CIFAR-10~\cite{cifar}, and ImageNet-1k~\cite{deng2009imagenet} (hereafter referred to as ImageNet for simplicity of exposition). 
We chose these five datasets because three (MNIST, FashionMNIST and SVHN) are compatible with the both our baselines \deepjanus~\cite{riccio2020deepjanus, 2023-Riccio-ICSE} and Sinvad~\cite{sinvad-tosem, kang2020sinvad}. This selection is also consistent with previous studies, such as Dola et al.~\cite{dola2024cit4dnn}. However, \tool can be applied to any image dataset. CIFAR-10 and ImageNet are used to demonstrate the generalizability of our approach to data sets where a model input representation is not available for \deepjanus and, thus, we compare against the Sinvad generative-based approach~\cite{sinvad-tosem} as a baseline. 

\head{MNIST} 
Dataset of handwritten digits~\cite{mnist} consisting of grayscale images $28\times28$ labeled with the corresponding digit (the possible classes range from 0 to 9). MNIST has 60,000 training inputs and 10,000 test inputs. As StyleGAN only allows square images of size $2^n$, we zero pad the images to scale them to size and duplicate channels resulting in an image of size $32\times32\times3$. 

\head{FashionMNIST}
Another dataset consisting of $28\times28$ grayscale images of Zalando's articles belonging to $10$ categories~\cite{fashionmnist}. The dataset has more complex patterns and variations than MNIST and contains $60,000$ images for training and $10,000$ for testing. Similarly to MNIST we again zero pad the data to make it compatible with StyleGAN having a shape of $32\times32\times3$. 

\head{SVHN}
A more complex dataset contains $32\times32\times3$ color digits of house numbers cropped from Google Street View images~\cite{svhn}. As the data is already compatible with StyleGANs no transformation were used.
It has $73,257$ training input and $26,032$ test input. 
The classification task is particularly challenging due to variations in lighting, background clutter, and the presence of distracting digits adjacent to the digit of interest. 

\head{CIFAR-10}
Another standard benchmark for image classification tasks is divided into $10$ classes of different objects~\cite{cifar} and divided into $50,000$ training images and $10,000$ testing images. Although the images are small ($32\times32\times3$), they contain visual complexities and variations of real-world objects, requiring models to extract meaningful features from low-resolution images. Again the data is compatible with StyleGAN as such no transformations were performed.

\head{ImageNet-1k} This dataset consists of over 14 million images spanning 1,000 classes. It has been widely recognized for its role in the ImageNet Large-Scale Visual Recognition Challenge (ILSVRC) since 2010, with the 2012 version being a benchmark standard for image classification tasks. Compared to the other three datasets, ImageNet-1k includes high-resolution images and a significantly broader range of categories. Due to the large size and the number of classes in ImageNet, we focused on the first ten classes, namely \textit{tench, goldfish, great white shark, tiger shark, hammerhead shark, electric ray, stingray, cock, hen, ostrich}. To make the data compatible with the StyleGAN used all images were transformed to fit into $128\times128\times3$.

\subsubsection{System Under Test}
We adopt a WideResNet-50-2 classifier~\cite{zagoruyko2016wide} available in the PyTorch library~\cite{paszke2017automatic} with pre-trained weights for ImageNet. The model achieves an accuracy of $0.816$ on ImageNet1k. For the other datasets (MNIST, FashionMNIST, SVHN, CIFAR-10), we train the same WideResNet-50-2 architecture with the default train data splits given in Torchvision~\cite{torchvision2016}.  We prioritized consistency across SUTs by using the same model architecture for all datasets, rather than using state-of-the-art or literature-sourced models, in order to reduce variability in our results due to architectural peculiarities.
The trained networks achieved an accuracy above 0.9 on the test split for all datasets, except for CIFAR-10, where the accuracy was 0.81.
For optimization in training, we use AdamW, which has demonstrated superior performance across multiple tasks due to its adaptive weight decay compared to traditional Adam~\cite{loshchilov2017decoupled}. Furthermore, we schedule our learning rates using the OneCycle strategy~\cite{smith2018superconvergence}, which initially increases the learning rate toward a maximum value and then decreases it again to a set minimum, forming a learning rate trajectory similar to a right-skewed Gaussian. This type of scheduling has been shown to improve convergence during training~\cite{smith2018superconvergence}.

\subsection{Metrics}

A boundary input is defined as an input that is close to the theoretical decision boundary. That is an input in which two or more classes are predicted as equally 
likely as described in (\ref{eq:eqd}). Note that a perfect equilibrium in prediction probabilies is unlikely due to the nature of floating point operations. Therefore the resulting input can either be failure-inducing, if it is misclassifed or it can be class preserving, if the class is predicted correctly. 
To address RQ\textsubscript{1}, we evaluated the quality of the generated boundary inputs using several metrics, described next.

\head{Boundary Distance ($\downarrow$)} In order to quantify whether a candidate is a good boundary input, we measure the Euclidean distance $d_2$ between the predicted classes $\hat{y}' \in \mathbb{R}^{||\mathcal{C}||}$ to the theoretical boundary \autoref{eq:boundary_dist}.
In this work, we specifically look at identifying boundary candidates between two classes rather than multiple, the theoretical boundary is assumed to be a vector $b \in \mathbb{R}^{||\mathcal{C}||}$ where $\sum b = 1$, and all non-zero elements are equal to $\frac{1}{1+||C_g||}$, that is in equilibrium between the classes used for boundary discovery $C_g$ and the original class $C_0$.The lower this measure, the better the boundary input.

\begin{equation}
    m_1(\hat{y}', b) = d_2(\hat{y}', b).\label{eq:boundary_dist}
\end{equation}

\head{Label Coverage ($\uparrow$) \& Escape Ratio ($\downarrow$)} Another important aspect of test case generation is the coverage of possible test outcomes. The Label Coverage indicates the distribution of the target labels, $\mathcal{Y}'_t = \{\hat{y}'_t \forall X'\}$, for the candidates of the boundaries, measured using the Kolmogorov–Smirnov distance \autoref{eq:coverage}. Here, $\mathcal{U}_t = \mathcal{U}\{\mathcal{C}\;\textbackslash\; \{c_0\}\}$ represents the uniform distribution of all possible target labels. A value of 1 indicates a uniform distribution in all possible classes, while a value of 0 indicates that all test cases share the same target label. Although achieving a perfect value of 1 is often infeasible due to the spatial separation of some classes on the decision surface, higher label coverage is generally preferred.

\begin{equation}
    m_3(\mathcal{U}_t, \mathcal{Y}_t) = d_{KS}(\mathcal{U}_t, \mathcal{Y}_t) .\label{eq:coverage}
\end{equation}

In addition to label coverage, the escape ratio quantifies the fraction of test cases that no longer consider the initial class as the origin of the boundary \autoref{eq:escape}. In this case, $\mathcal{Y}$ represents the set of predictions on all initial seeds, $\mathcal{Y}'$ is the set of predictions for the candidates generated, and $\left[\cdot\right]$ is the Iverson bracket. The subscript on the predictions denotes the  indices taken when an argmax is applied to the vector. This measure is critical because, when testing boundaries for a specific class, we are interested only in boundaries that relate to the original class. Thus, the escape ratio should be small to ensure that the generated candidates remain relevant to the objective.

\begin{equation}
    m_4(\mathcal{P}, \mathcal{Y}, \mathcal{Y}') = \frac{1}{||\mathcal{P}||}\sum_{\hat{y} \in \mathcal{Y}, \hat{y}' \in \mathcal{Y}'} [\hat{y}_{1st} \notin  \{\hat{y}'_{1st}, \hat{y}'_{2nd} \}].\label{eq:escape}
\end{equation}

\head{Laplacian Variance of Image Differences ($\uparrow$)} This measure aims to quantify the change in information between the boundary input and the initial input \autoref{eq:lap_var}. Essentially, it indicates whether the method produces functionally different outputs or merely corrupts or blurs an image, common phenomena observed when applying simple pertubations to the latent space.
Here, $L$ is a $3\times3$ Laplacian kernel, applied using convolution on the differences of two images. The higher this measure, the better the boundary input. 

\begin{equation}\label{eq:lap_var}
    m_2(X, X') = Var((X-X')\ast L).
\end{equation}

About RQ\textsubscript{2}, we evaluate the performance of \tool by computing  the time required, in seconds, to generate a single candidate solution. For comparison we aggregate the runtime across datasets for each method, and report the mean runtime and its standard deviation. 

Concerning RQ\textsubscript{3}, we performed an evaluation study with human assessors to evaluate the quality of the generated inputs. Quality was assessed with several characteristics, described as follows.
Label preservation describes whether the original class label is still assigned to the generated candidate by the evaluator. The inverse of which is target preservation, which quantifies if the boundary target is visually depicted in the generated image. When combining these two we get boundary preservation, as the generated image shows visual elements of either classes of the boundary. The latter is simply the sum of the first two. Finally, we measure the validity~\cite{2023-Riccio-ICSE}, i.e., whether any class within the considered domain was associated by the evaluator, meaning the image still has valid syntactic features to the human observer.

Answering RQ\textsubscript{4}, we investigate how the latent space is used by \tool in the implementation used for the experiments. Additionally we compare this usage to a configuration with a different selection of objective functions $\{\omega_{dcb}\}$, only evaluating dynamic confidence balance and $\{ \omega_{dcb}, \omega_{d2}\}$, which includes archive sparsity for novelty of solutions. 
The evaluation is done by aggregating the genome weights, as they dictate the usage of the latent vectors. Specifically, we look at the distribution of those weights in combination with general uniformity of the distributions to showcase differences in the extents of manipulation.

\subsection{Baseline Approaches} 
To assess the relevance of our approach, we compare \tool against Sinvad and \deepjanus, two state-of-the-art test generator for the exploration of the frontier of behaviors of DL systems. 

Sinvad~\cite{sinvad-tosem} is a latent manipulation-based tool that leverages Variational Autoencoders (VAEs) as its generative networks. Specifically, Sinvad encodes test inputs into latent vectors using a VAE trained on the corresponding training set. Since the approach relies on population-based optimization, the initial population is derived from the latent vector of the original images, perturbed with random noise to simulate a diverse set of slightly altered inputs. Sinvad then optimizes toward a fitness function through uniform crossover of latent vectors and mutation via noise. The mutation severity is dynamically adjusted: if progress toward the objective function stalls, the mutation magnitude decreases accordingly. In our study we compare Sinvad against \tool on all datasets using the pre-trained VAEs available in the replication package~\cite{sinvad-tosem}.

\deepjanus~\cite{riccio2020deepjanus} is a representative of model-based approaches and uses a multi-objective search-based algorithm to mutate the control points of a model of the inputs, to generate pairs of inputs that are close to each other, yet produce different behaviors of the DL system~\cite{riccio2020deepjanus}. 
The input model representation is obtained through a vectorization operation, which produces a sequence of control points that are iteratively displaced to achieve slight modifications. The input image can then be reconstructed through a rasterization operation.
Our comparison focused on the MNIST, FashionMNIST, and SVHN datasets, as \deepjanus's model representation supports these benchmarks. For CIFAR-10 and ImageNet, \deepjanus is not applicable since an appropriate input model is not available for such a feature-rich dataset, and cannot be created with the adopted vectorization-rasterization approach, as noted by its authors~\cite{riccio2020deepjanus,2023-Riccio-ICSE}. 

\subsection{Procedure}\label{sec:exp_set}

Our approach requires generating seeds by sampling latent vectors using a conditional StyleGAN architecture. For MNIST, FashionMNIST, and SVHN, pre-trained conditional StyleGAN2 networks are not available. Thus, 
we trained them on the training partition of each dataset, following the training configurations and guidelines of the original paper~\cite{DBLP:journals/corr/abs-1912-04958}. 
To monitor the model's performance during training, we used the Fr{\'e}chet Inception Distance (FID) metric~\cite{DBLP:journals/corr/HeuselRUNKH17}.
The final FID score obtained is 0.91 for MNIST, 2.34 for FashionMNIST and 4.20 for SVHN, which is in line with the original paper~\cite{DBLP:journals/corr/abs-1912-04958}.
For CIFAR-10, we used pre-trained StyleGAN2 networks available in the literature~\cite{DBLP:journals/corr/abs-2006-06676}. For ImageNet, we used a pre-trained StyleGAN-XL~\cite{sauer2022styleganXL}.

For RQ\textsubscript{1}, for all applicable datasets, we execute all test generators (\tool, Sinvad, \deepjanus) using a budget of $15,000$ predictions of the SUT per boundary candidate to be optimized for. 
Given the varying approaches of the methods, we define the SUT as the determinant of the budget, ensuring consistency across experiments. While the budget may be reached, it does not need to be fully utilized, as some methods may terminate earlier. Particularly, we instruct the tools to search for 10 boundary candidates for each class, giving us a total budget of $1.5M$ predictions per test method and dataset. Overall, our study includes nearly $20M$ predictions ($2$ tools, \tool and Sinvad $\times$ $1.5M$ predictions $\times$ $5$ datasets + $1.5M$ $\times$ $3$ datasets for \deepjanus).

For RQ\textsubscript{2}, in addition to the number of iterations used, we also record the runtime, acknowledging that implementation efficiency can influence performance.

For RQ\textsubscript{3}, as image metrics often do not coincide with human perception~\cite{lambertenghi2024assessing}, we use a human evaluation study to quantify the quality of generated cases. The evaluators are recruited on AWS Mechanical Turk, with an attention question incorporated to filter out inattentive or disengaged responses~\cite{sorokin2008utility}. The attention question was given with the datasets reference images, asking the participants to select a specific class.
Prior to the evaluation, images of the original datasets were shown to make the assessors familiar with the classes. For each dataset, we randomly selected 10 generated images from each method (\tool, Sinvad, DeepJanus), covering all classes. We ask the participants to select all classes they can recognize in the provided image, with an ``Not applicable'' option if they could not identify any of the provided classes. Each participant was only shown samples from a single dataset. We recruited 30 distinct evaluators for each dataset, with $\sim 20$ valid respondents per dataset, as some had to be discarded due to incorrect attention question answers.

For RQ\textsubscript{4}, we monitor the final latent interpolation weights, found in optimization. As the initial strategies are all initialized randomly, it is interesting to see whether certain types of genomes are more likely to appear at the end of optimization. For each found candidate solution, we therefore have a corresponding latent interpolation weight vector which is then used for the analysis.

\subsection{Results}\label{sec:results}
\subsubsection{RQ\textsubscript{1} (effectiveness)}\label{sec:rq1}

\begin{figure}[t]
    \centering
    \begin{subfigure}[b]{0.32\textwidth}
        \centering
        \includegraphics[width=\textwidth]{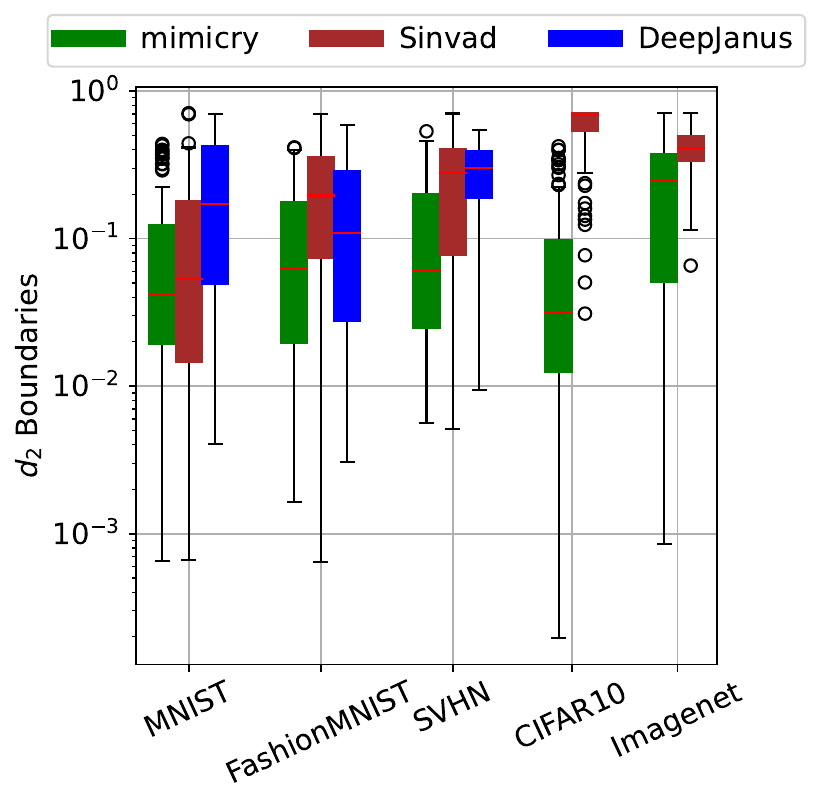}
        \caption{Boundary Distance.}
        \label{fig:boundary_comp}
    \end{subfigure}
    \begin{subfigure}[b]{0.315\textwidth}
        \centering
        \includegraphics[width=\textwidth]{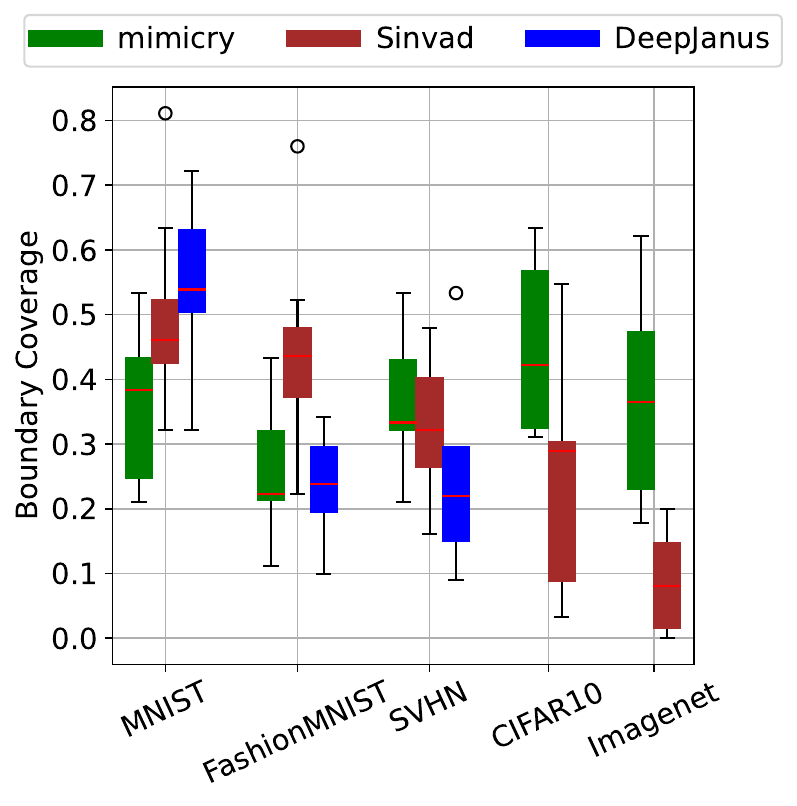}
        \caption{Label Coverage.}
        \label{fig:lcer}
    \end{subfigure}
    \begin{subfigure}[b]{0.32\textwidth}
        \centering
        \includegraphics[width=\textwidth]{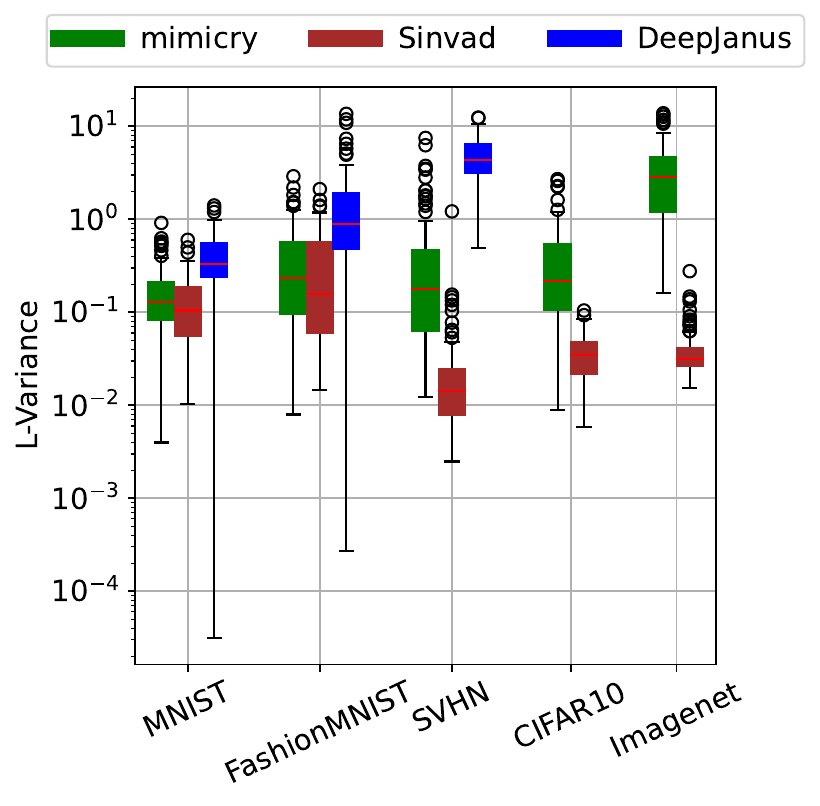}
        \caption{Laplacian Variance.}
        \label{fig:lap_var}
    \end{subfigure}
    \caption{Effectiveness results (RQ\textsubscript{1}).}
     \label{fig:rq1}
\end{figure}

\autoref{fig:rq1} reports two plots related to the considered effectiveness metrics. Each plot displays the distribution of the metrics as boxplots, aggregated across all datasets.

Considering boundary distance (\autoref{fig:boundary_comp}), \tool consistently generates boundary inputs characterized by lower boundary distances for all datasets, showing the competitiveness of our approach at generating inputs close to the equilibrium between the source and target class. Related to the baselines, \deepjanus's inputs exhibits higher distance, arguably due to the model-based transformation, which makes it impossible to perform fine-grained input manipulations. Concerning Sinvad, it exhibits competitive scores for simple datasets (MNIST, FashionMNIST, and SVHN), even tough worse than those of \tool. In contrast, for more complex datasets such as CIFAR-10 and ImageNet, the effectiveness of Sinvad is particularly low, especially for CIFAR-10. 

\begin{table}[t]
\caption{RQ\textsubscript{1}: Escape ratio for all approaches and datasets.}
\label{tab:escr}
\renewcommand{\arraystretch}{1.1}
\begin{tabular}{lccccc}
\toprule
          & \textbf{MNIST} & \textbf{FashionMNIST} & \textbf{SVHN} & \textbf{CIFAR-10} & \textbf{ImageNet} \\ 
          
\midrule

\tool      & \textbf{0}              & \textbf{0}                     & \textbf{0.01}          & \textbf{0}                & \textbf{0.07}             \\
Sinvad    & 0.01           & 0.05                  & 0.23          & 0.30              & 0.59              \\
DeepJanus & 0.02           &  0.13   & 0.28          &   N/A   &  N/A \\
\bottomrule
\end{tabular}
\end{table}

Regarding label coverage (\autoref{fig:lcer}), DeepJanus outperforms GenAI-based methods in label coverage for MNIST. However, this trend reverses for SVHN. With more complex datasets, \tool outperforms Sinvad in terms of label coverage. It is important to note that label coverage alone does not provide a complete picture. For example, applying noise to images may result in high label coverage because of a wide variation in target labels. However, for a boundary candidate to be useful, it must remain relevant as a boundary candidate between the original class and others. This is exactly what the escape ratio quantifies. \autoref{tab:escr} reports the average escape ratio across all datasets, which shows that \tool outperforms the competing methods methods across all datasets, indicating that the generated boundary inputs are more likely to be useful for testing specific boundaries.

\begin{figure}[b]
    \centering
    \begin{subfigure}[b]{0.32\textwidth}
        \centering
        \includegraphics[width=\textwidth]{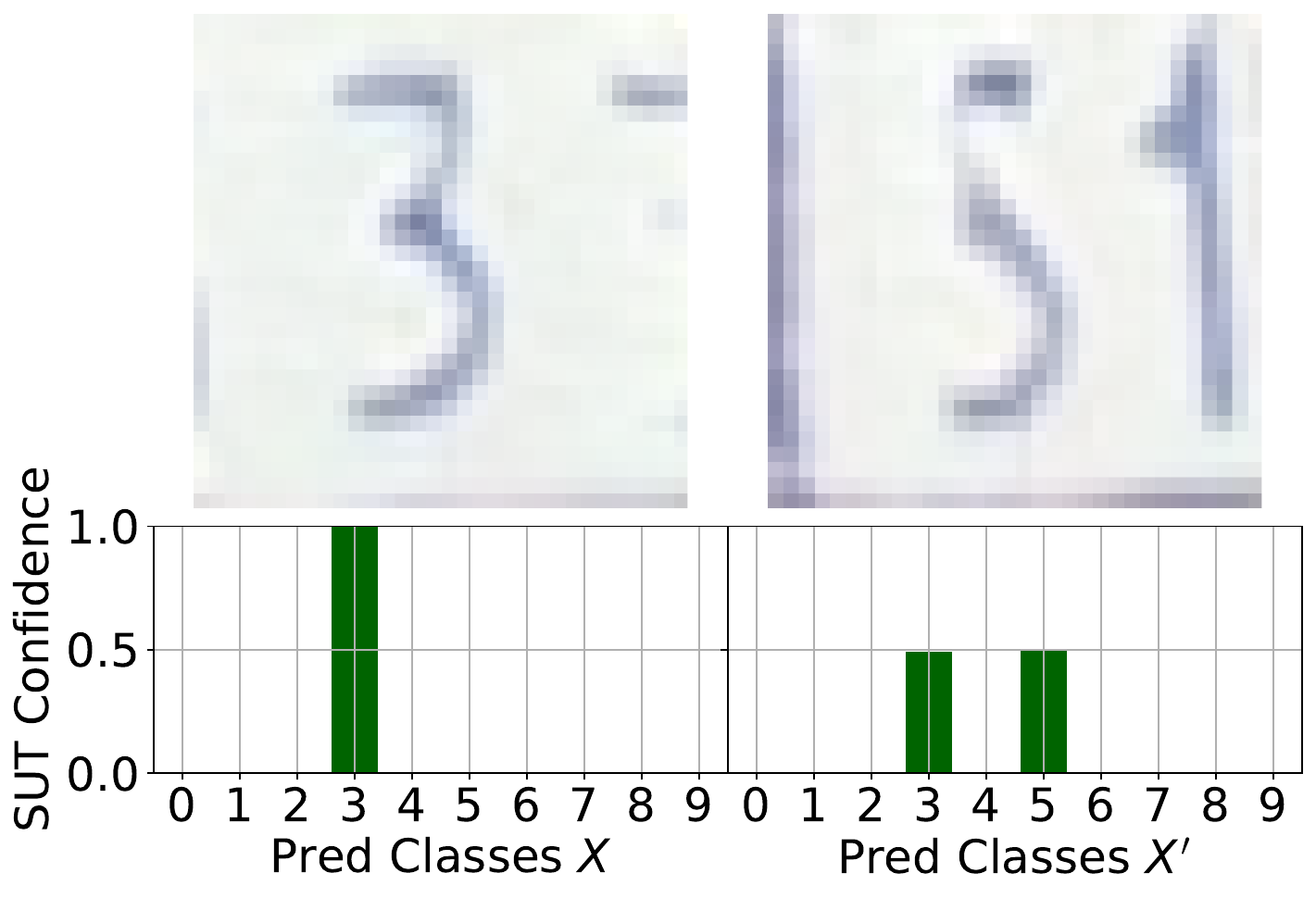}
        \caption{3 $\rightsquigarrow$ 5 by \tool.}
    \end{subfigure}%
    \begin{subfigure}[b]{0.32\textwidth}
        \centering
        \includegraphics[width=\textwidth]{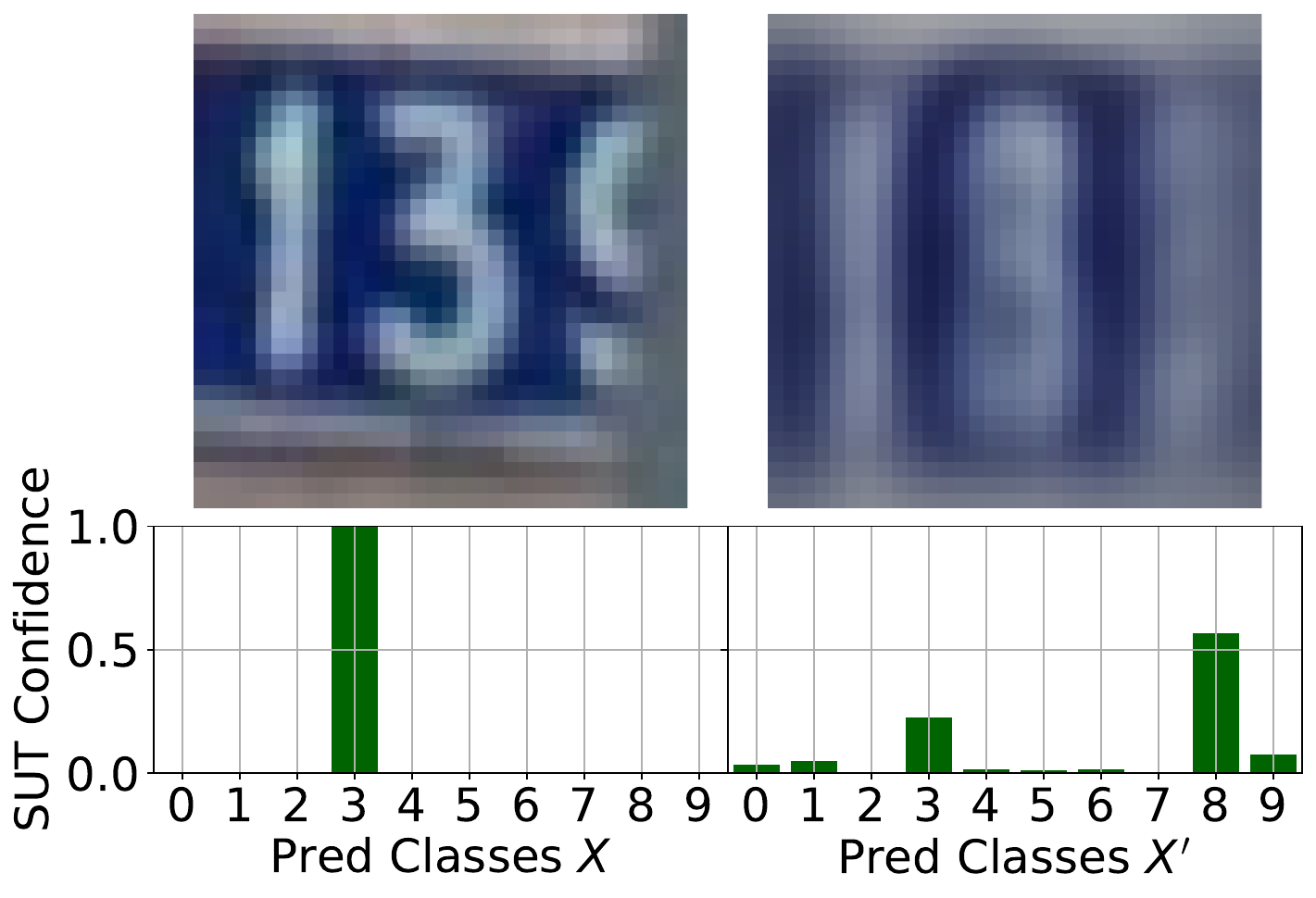}
        \caption{3 $\rightsquigarrow$ 8 by Sinvad.}
    \end{subfigure}%
    \begin{subfigure}[b]{0.32\textwidth}
        \centering
        \includegraphics[width=\textwidth]{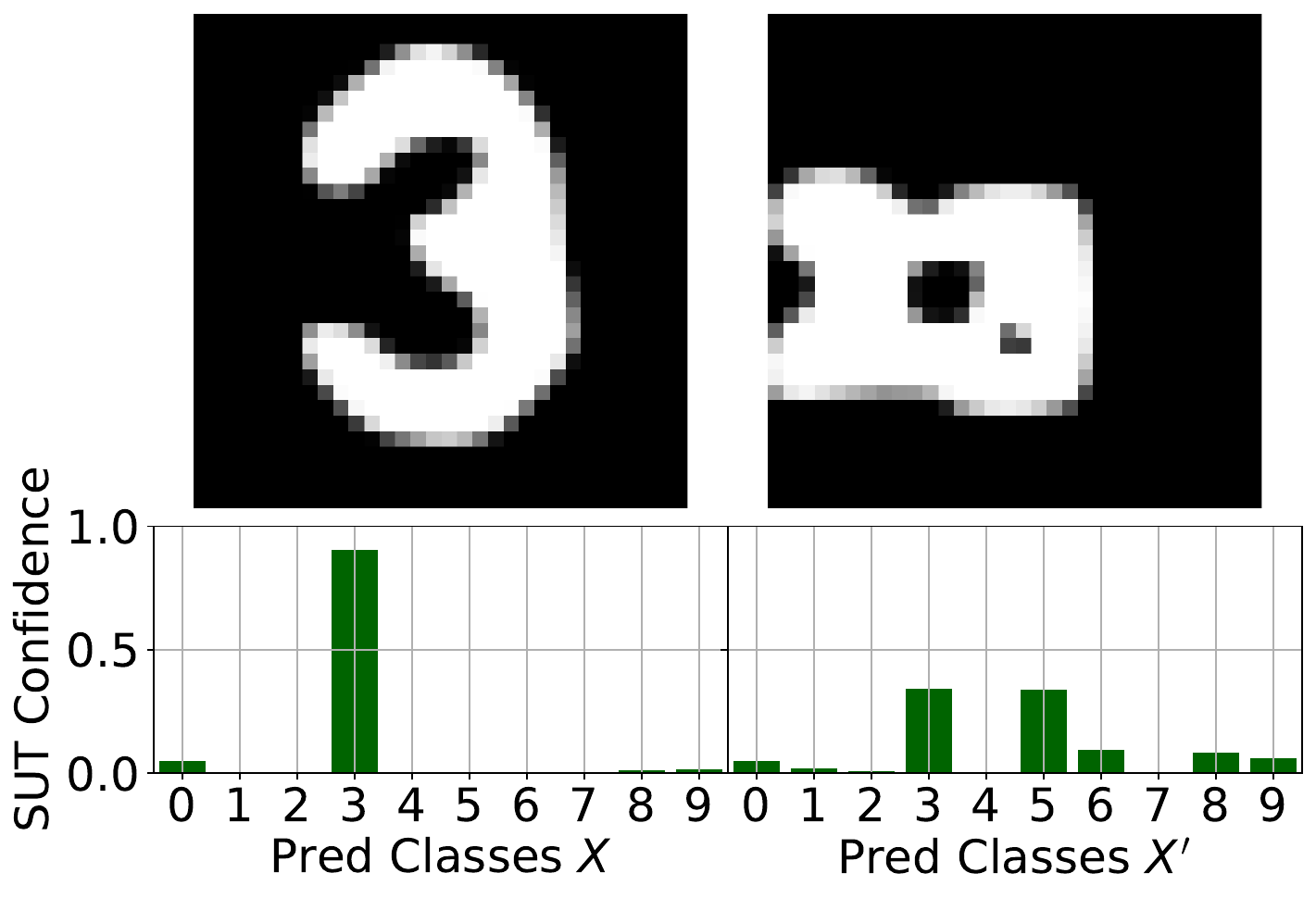}
        \caption{3 $\rightsquigarrow$ 5 by DeepJanus.}
        \label{fig:dj_svhn}
    \end{subfigure}
    \vspace{0.5em}
    \begin{subfigure}[b]{0.49\textwidth}
        \centering
        \includegraphics[width=\textwidth]{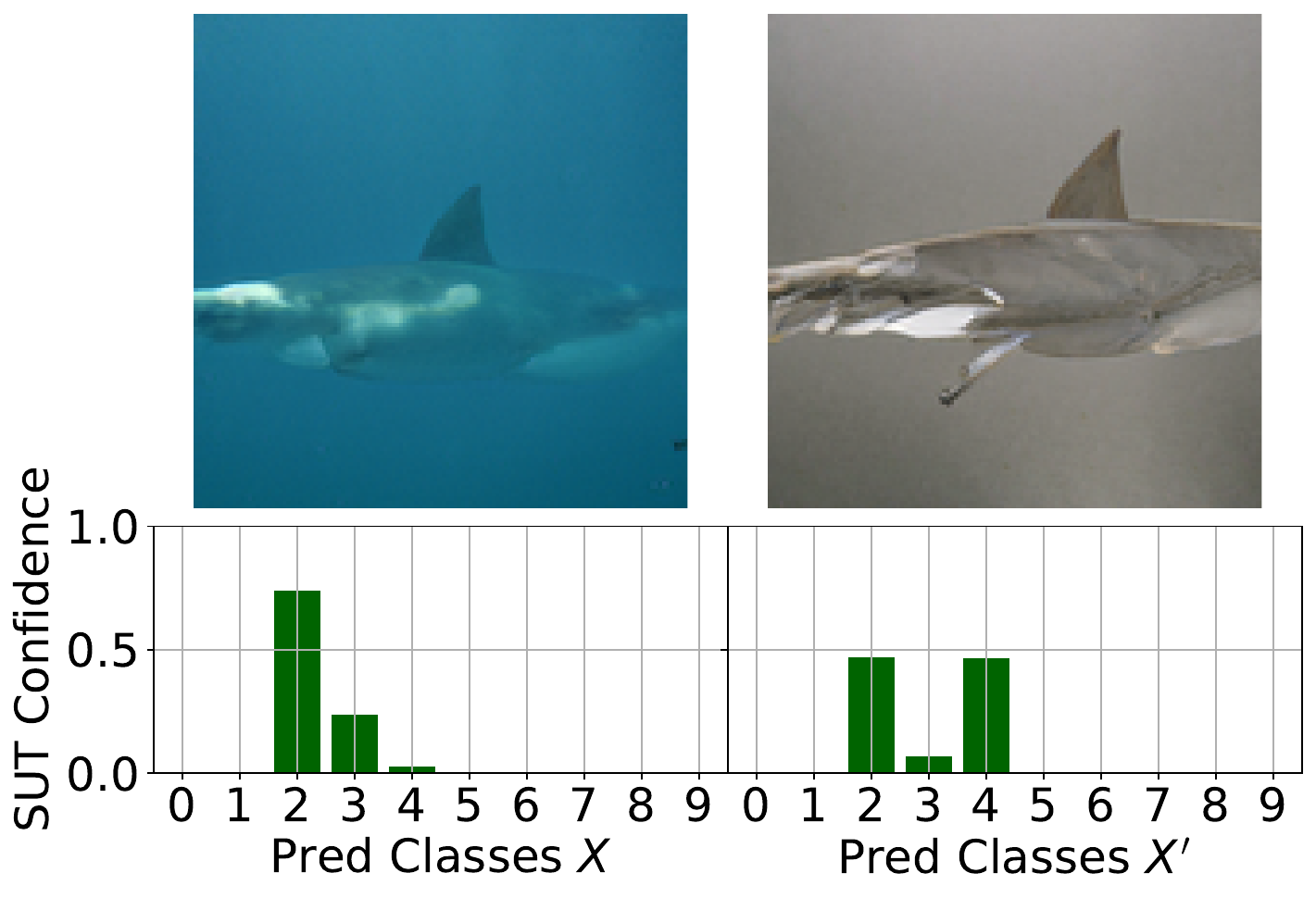}
        \caption{\textit{white shark} $\rightsquigarrow$ \textit{hammerhead shark} by \tool.}
    \end{subfigure}%
    \begin{subfigure}[b]{0.49\textwidth}
        \centering
        \includegraphics[width=\textwidth]{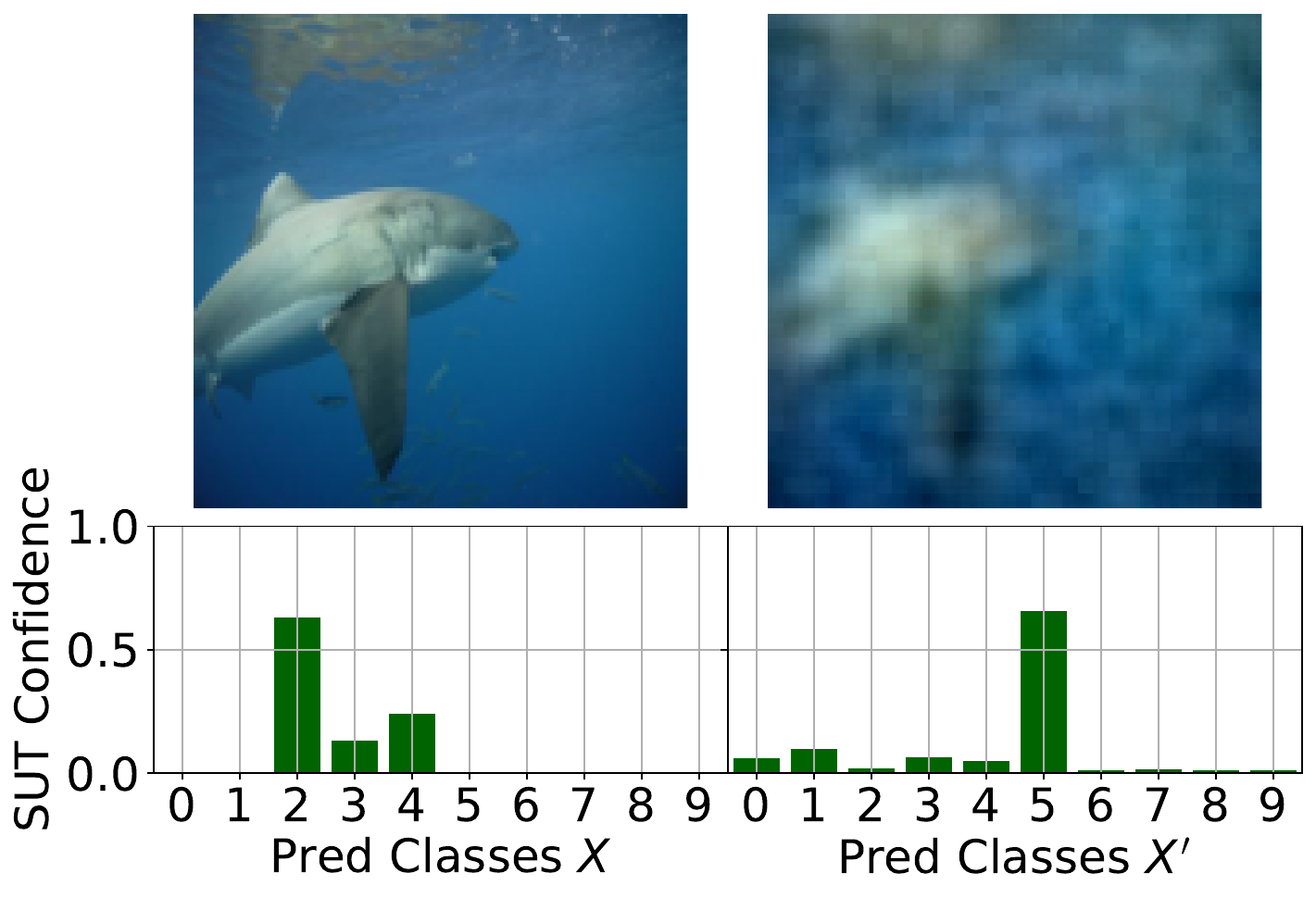}
        \caption{\textit{white shark} $\rightsquigarrow$ \textit{electric ray} by Sinvad.}\label{fig:comp_e}
    \end{subfigure}%
    \caption{\textbf{SVHN \& ImageNet}: Original image and corresponding boundary input with SUT confidence.}\label{fig:comp}
\end{figure}

Regarding Laplacian variance, \autoref{fig:lap_var} highlights that \deepjanus performs better when a model is available. However, this metric is skewed due to the way \deepjanus generates solutions by modifying vector paths, which results in pixels being either black or white (see \autoref{fig:dj_svhn}). This artificially increases variance, as the Laplacian filter responds strongly to sharp edges.
In contrast, generative-based solutions produce values across the entire spectrum, leading to smoother transitions and less pronounced edges.
When comparing \tool and Sinvad, an interesting trend emerges, consistent with previous metrics. As data complexity increases, Sinvad's performance drops significantly, producing images that appear blurred rather than functionally manipulated (see \autoref{fig:comp_e}).

\begin{table}[t]
\caption{RQ\textsubscript{1}: Statistical analysis. Significant $p$-values are boldfaced.}
\label{tab:stat_comp}
\renewcommand{\arraystretch}{1.1}
\resizebox{\columnwidth}{!}{
\begin{tabular}{l
cccccccc
}

\toprule

& \multicolumn{2}{c}{\textbf{MNIST}} & \multicolumn{2}{c}{\textbf{FashionMNIST}} & \multicolumn{2}{c}{\textbf{SVHN}} & \textbf{CIFAR-10} & \textbf{ImageNet} \\

\cmidrule(r){2-3}
\cmidrule(r){4-5}
\cmidrule(r){6-7}
\cmidrule(r){8-8}
\cmidrule(r){9-9}

  &  Sinvad  & DeepJanus  &  Sinvad  & DeepJanus    & Sinvad & DeepJanus        &  Sinvad          & Sinvad          \\ 

\midrule
 
Boundary distance& 0.283 & \textbf{1.19e-7\textcolor{orange}{$\bullet$}} & \textbf{9.51e-7\textcolor{orange}{$\bullet$}}& 
\textbf{0.005\textcolor{darkgreen}{$\bullet$}} &
\textbf{5.21e-9 \textcolor{orange}{$\bullet$}} & \textbf{7.7e-15 \textcolor{darkgreen}{$\bullet$}}  &  \textbf{4.08e-31 \textcolor{darkgreen}{$\bullet$}} & \textbf{1.16e-11 \textcolor{darkgreen}{$\bullet$}}\\
Laplacian variance& \textbf{0.014 \textcolor{red}{$\bullet$}} & $\sim$ 1 & 0.083 & $\sim1$ & \textbf{1.61e-27 \textcolor{orange}{$\bullet$}} & $\sim$ 1 & \textbf{1.03e-25 \textcolor{darkgreen}{$\bullet$}} & \textbf{1.31e-34 \textcolor{darkgreen}{$\bullet$}} \\
Label coverage& 0.986  & 0.998 & 0.998 &0.32& 0.213 & \textbf{0.006 \textcolor{darkgreen}{$\bullet$}}  & \textbf{0.001\textcolor{darkgreen}{$\bullet$}}& \textbf{1.41e-40 \textcolor{darkgreen}{$\bullet$}}\\ 

\bottomrule

\end{tabular}
}
\end{table}

Aggregating these measures, we are interested in whether these differences have statistical significance. Therefore we employ a one-tailed Mann–Whitney U test~\cite{Wilcoxon1945} (with $\alpha = 0.05$) between \tool and the baseline methods. Additionally, we calculate the Cohen's $d$ effect size~\cite{cohen1988statistical}, whose magnitude is indicated in \autoref{tab:stat_comp} by a colored bullet ($\bullet$), where $\textcolor{darkgreen}\bullet = d > 1$ (large) $,\;  \textcolor{orange}\bullet = 1 \geq d > 0.5$ (medium) $,\; \textcolor{red}\bullet = d \leq 0.5$ (small). The statistical results confirm the trend observed in the figures, where \tool performs well on all datasets and outperforms the baselines especially as their complexity increases. 

\begin{tcolorbox}[boxrule=0pt,sharp corners,boxsep=2pt,left=2pt,right=2pt,top=2.5pt,bottom=2pt]
\begin{center}
\begin{minipage}[t]{0.99\linewidth}
\textbf{RQ\textsubscript{1} (effectiveness)}: \textit{
\tool significantly outperforms baseline methods in boundary distance, label coverage, and escape ratio. As dataset complexity increases, it generates more relevant and effective candidates for boundary testing, demonstrating a clear advantage over existing approaches in manipulating image content for functional testing.
}
\end{minipage}
\end{center}
\end{tcolorbox}

\subsubsection{RQ\textsubscript{2} (efficiency)}\label{sec:rq2}

\begin{table}[b]
\caption{Mean runtime \& standard deviation per 15,000 iterations (in seconds) and trainable parameters.} 
\renewcommand{\arraystretch}{1.1}
\begin{tabular}{l
ccccc
}
\toprule

& \textbf{MNIST} & \textbf{FashionMNIST} & \textbf{SVHN}  & \textbf{CIFAR-10} & \textbf{ImageNet}\\ 
          
\midrule

\tool      & 78.98 $\pm$ 2.03 & 76.56 $\pm$ 1.02 & 77.97 $\pm$ 1.41 & 81.04 $\pm$ 1.71 & 412.56 $\pm$ 14.37 \\ 
\textit{\Small $\phi_G$ params}& \textit{\Small 21M}&\textit{\Small 21M}&\textit{\Small 21M}&\textit{\Small 20M}&\textit{\Small 158M} \\ 
[0.5em]

Sinvad    & \textbf{3.34 $\pm$ 0.36} & \textbf{3.36 $\pm$ 0.45} & \textbf{3.54 $\pm$ 0.49 } & \textbf{4.91 $\pm$ 1.46} & \textbf{7.77 $\pm$ 1.72} \\ 
\textit{\Small $\phi_G$ params}& \textit{\Small 4M} & \textit{\Small 4M}  & \textit{\Small 13M} & \textit{\Small 83M} & \textit{\Small 79M} \\ [0.5em]

DeepJanus & 3.90 $\pm$ 0.26  & 7.85 $\pm$ 22.45 & 105.32 $\pm$ 257.80 & N/A & N/A              \\ 

\bottomrule

\end{tabular}\label{tab:runt}
\end{table}

\autoref{tab:runt} shows the efficiency results, normalized to $15,000$ iterations to make the methods comparable. Sinvad proved to be the fastest approach, outperforming \tool and DeepJanus in terms of raw execution time. 
An interesting characteristic of \tool is the relative stability of its mean runtime, as evidenced by the corresponding standard deviation. Notably, all StyleGAN2-based solutions (MNIST and CIFAR-10) exhibit a relatively consistent runtime, whereas the StyleGAN-XL-based ImageNet generator incurs higher computational costs. This is due to the number of trainable parameters being similar or equal in the StyleGAN2 cases (\autoref{tab:runt}). In contrast, Sinvad employs an early termination condition, resulting in a relatively large standard deviation compared to its mean runtime. This condition reveals that as data complexity increases, Sinvad does no longer control the generated candidates, leading to an insufficient usage of computational budget. When applied to CIFAR-10 and ImageNet, Sinvad terminates at the minimal possible budget used due to an internal mechanism that reduces mutation size when conditions remain unsatisfied. This reduction ultimately triggers early termination (more details are available in our appendix).

The results for DeepJanus reveal a significant increase in both mean runtime and standard deviation as dataset complexity increases. This behavior is an artifact of the two mutation operators employed, which depend on the presence of specific patterns in the input's SVG paths. When these patterns are absent, the mutation operations are ineffective and lead to prolonged computation.

\begin{tcolorbox}[boxrule=0pt,sharp corners,boxsep=2pt,left=2pt,right=2pt,top=2.5pt,bottom=2pt]
\begin{center}
\begin{minipage}[t]{0.99\linewidth}
\textbf{RQ\textsubscript{2} (efficiency)}: \textit{
\tool maintains a consistent runtime, unlike Sinvad, which exhibits high variability due to early termination, and DeepJanus, which slows down as dataset complexity increases. Although \tool is significantly slower than Sinvad, its effectiveness results (RQ\textsubscript{1}) combined with efficiency demonstrate a superior trade-off between quality and speed.
}
\end{minipage}
\end{center}
\end{tcolorbox}

\subsubsection{RQ\textsubscript{3} (quality)}\label{sec:rq3}

\autoref{tab:hum} shows the results of the human study. The table reports, for each dataset and approach, the average label preservation and target preservation scores, as well as the boundary preservation and validity scores.
\tool outperforms all baselines in terms of validity because the generated images are more likely to have a visibly recognizable class for the human observers. When looking at the boundary preservation a similar trend emerges, with the exception of ImageNet, in which Sinvad scores the best results.

For label and target preservation, \tool outperforms the baselines in most datasets, with some exceptions being \deepjanus in SVHN and Sinvad in CIFAR10 and ImageNet. The ImageNet results are especially interesting as they have implications for human studies when doing boundary testing, which seems to be challenging in more complex and feature-rich datasets. 

\begin{table}[t]
\caption{Human Image Evaluation Statistics}\label{tab:hum}
\resizebox{\columnwidth}{!}{
\begin{tabular}{llllll}
\toprule    &    & \textbf{Label Preservation} & \textbf{Target Preservation} & \textbf{Boundary Preservation} & \textbf{Validity} \\ \midrule
\rowcolor[HTML]{FFFFFF} 
\cellcolor[HTML]{FFFFFF}                                    & \multicolumn{1}{l}{\cellcolor[HTML]{FFFFFF}\tool}      & \textbf{0.460}            & \textbf{0.422}             & \textbf{0.882}               & \textbf{0.965} \\
\rowcolor[HTML]{FFFFFF} 
\cellcolor[HTML]{FFFFFF}                                    & \multicolumn{1}{l}{\cellcolor[HTML]{FFFFFF}Sinvad}    & 0.360                     & 0.268                      & 0.628                        & 0.790          \\
\rowcolor[HTML]{FFFFFF} 
\multirow{-3}{*}{\cellcolor[HTML]{FFFFFF}\textit{MNIST}}    & \multicolumn{1}{l}{\cellcolor[HTML]{FFFFFF}DeepJanus} & 0.690                     & 0.090                      & 0.780                        & 0.830          \\ \midrule
                                                            & \multicolumn{1}{l}{\tool}                              & \textbf{0.417}            & \textbf{0.280}             & \textbf{0.697}               & \textbf{0.944} \\
                                                            & \multicolumn{1}{l}{Sinvad}                            & 0.272                     & 0.210                      & 0.481                        & 0.757          \\
\multirow{-3}{*}{FashionMNIST}                     & \multicolumn{1}{l}{DeepJanus}                         & 0.266                     & 0.207                      & 0.474                        & 0.816          \\ \midrule
\rowcolor[HTML]{FFFFFF} 
\cellcolor[HTML]{FFFFFF}                                    & \multicolumn{1}{l}{\cellcolor[HTML]{FFFFFF}\tool}      & 0.240                     & \textbf{0.325}             & \textbf{0.565}               & \textbf{0.865} \\
\rowcolor[HTML]{FFFFFF} 
\cellcolor[HTML]{FFFFFF}                                    & \multicolumn{1}{l}{\cellcolor[HTML]{FFFFFF}Sinvad}    & 0.134                     & 0.233                      & 0.366                        & 0.673          \\
\rowcolor[HTML]{FFFFFF} 
\multirow{-3}{*}{\cellcolor[HTML]{FFFFFF}SVHN}     & \multicolumn{1}{l}{\cellcolor[HTML]{FFFFFF}DeepJanus} & \textbf{0.328}            & 0.078                      & 0.405                        & 0.790          \\ \midrule
                                                            & \multicolumn{1}{l}{\tool}                              & \textbf{0.470}            & 0.089                      & \textbf{0.559}               & \textbf{0.803} \\
\multirow{-2}{*}{CIFAR-10}                          & \multicolumn{1}{l}{Sinvad}                            & 0.255                     & \textbf{0.120}             & 0.375                        & 0.589          \\ \midrule
\rowcolor[HTML]{FFFFFF} 
\cellcolor[HTML]{FFFFFF}                                    & \multicolumn{1}{l}{\cellcolor[HTML]{FFFFFF}\tool}      & 0.188                    & \textbf{0.064}                      & 0.252                        & \textbf{0.711} \\
\rowcolor[HTML]{FFFFFF} 
\multirow{-2}{*}{\cellcolor[HTML]{FFFFFF}ImageNet} & \multicolumn{1}{l}{\cellcolor[HTML]{FFFFFF}Sinvad}    & \textbf{0.243}            & \textbf{0.064}             & \textbf{0.307}               & 0.707          \\ \bottomrule
\end{tabular}
}
\end{table}

\begin{tcolorbox}[boxrule=0pt,sharp corners,boxsep=2pt,left=2pt,right=2pt,top=2.5pt,bottom=2pt]
\begin{center}
\begin{minipage}[t]{0.99\linewidth}
\textbf{RQ\textsubscript{3} (quality)}: \textit{
\tool outperforms the baselines across datasets except for ImageNet, where Sinvad had higher preservation scores but lower validity. This suggests that while \tool is generally effective in maintaining both validity and label preservation, its optimization for DL decision boundaries may reduce interpretability in complex datasets.
}
\end{minipage}
\end{center}
\end{tcolorbox}

\subsubsection{RQ\textsubscript{4} (latent space usage)}\label{sec:rq4}

\begin{figure}[htbp]
    \centering
    \begin{subfigure}[b]{0.49\textwidth}
        \centering
        \includegraphics[width=\textwidth]{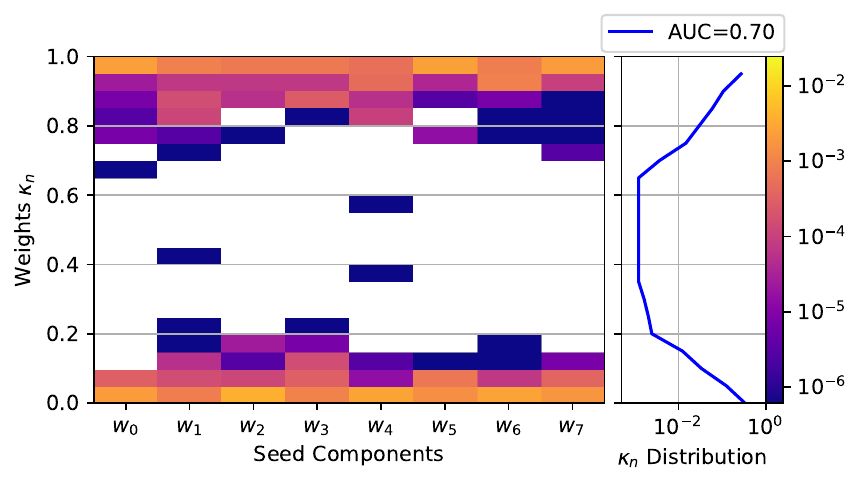}
        \vspace{-2em}
        \caption{MNIST}
    \end{subfigure}%
    \begin{subfigure}[b]{0.49\textwidth}
        \centering
        \includegraphics[width=\textwidth]{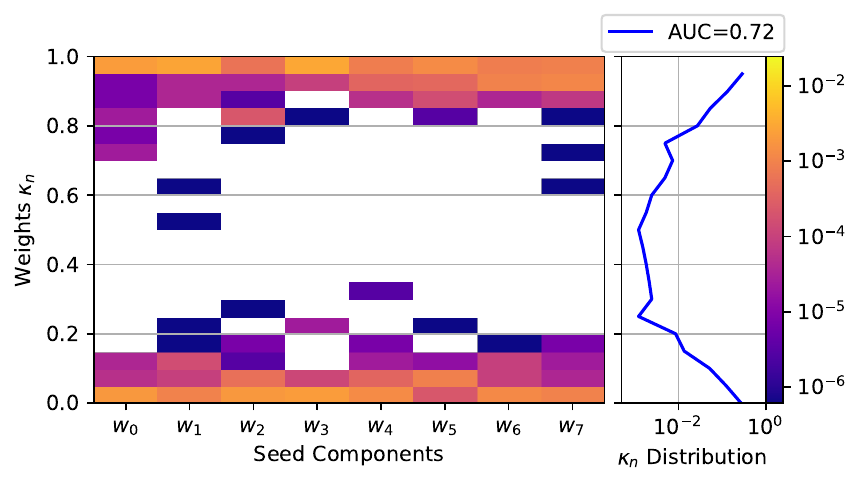}
        \vspace{-2em}
        \caption{FashionMNIST}
    \end{subfigure}%
    \vspace{0.5em}
    \begin{subfigure}[b]{0.49\textwidth}
        \centering
        \includegraphics[width=\textwidth]{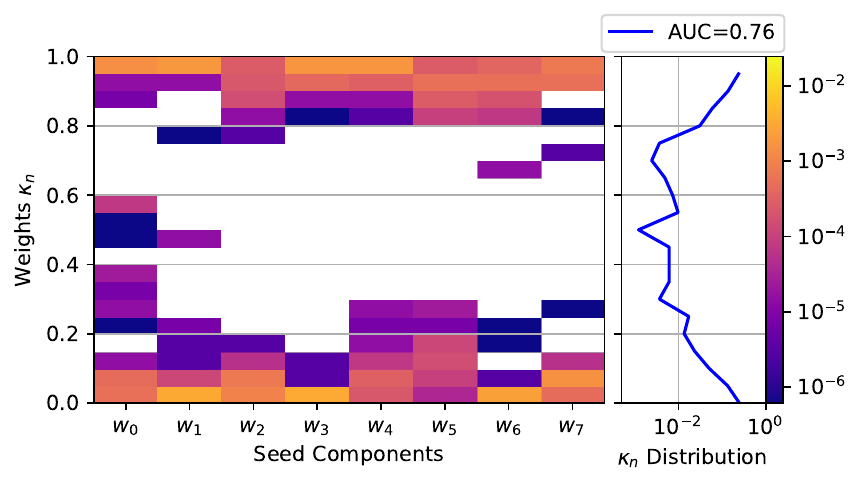}
        \vspace{-2em}
        \caption{SVHN}
    \end{subfigure}%
    \begin{subfigure}[b]{0.49\textwidth}
        \centering
        \includegraphics[width=\textwidth]{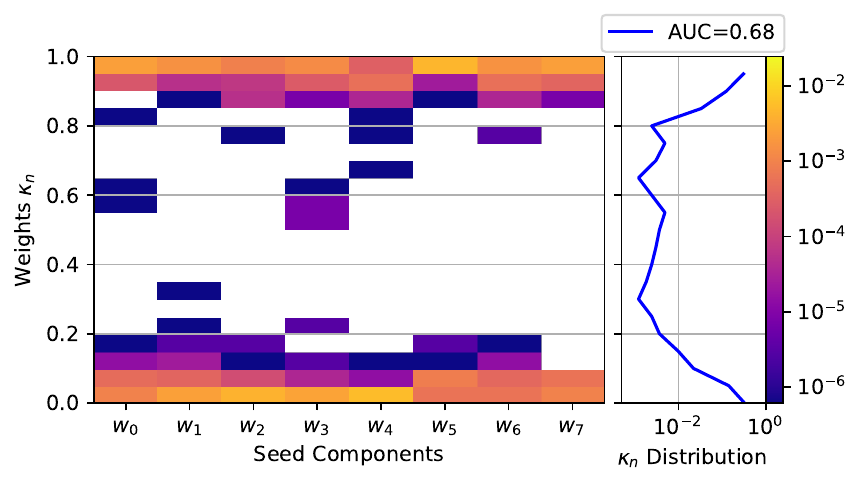}
        \vspace{-2em}
        \caption{CIFAR10}
        \label{fig:heat_d}
    \end{subfigure}%
    \vspace{0.5em}
    \begin{subfigure}[b]{0.7\textwidth}
        \centering
        \includegraphics[width=\textwidth]{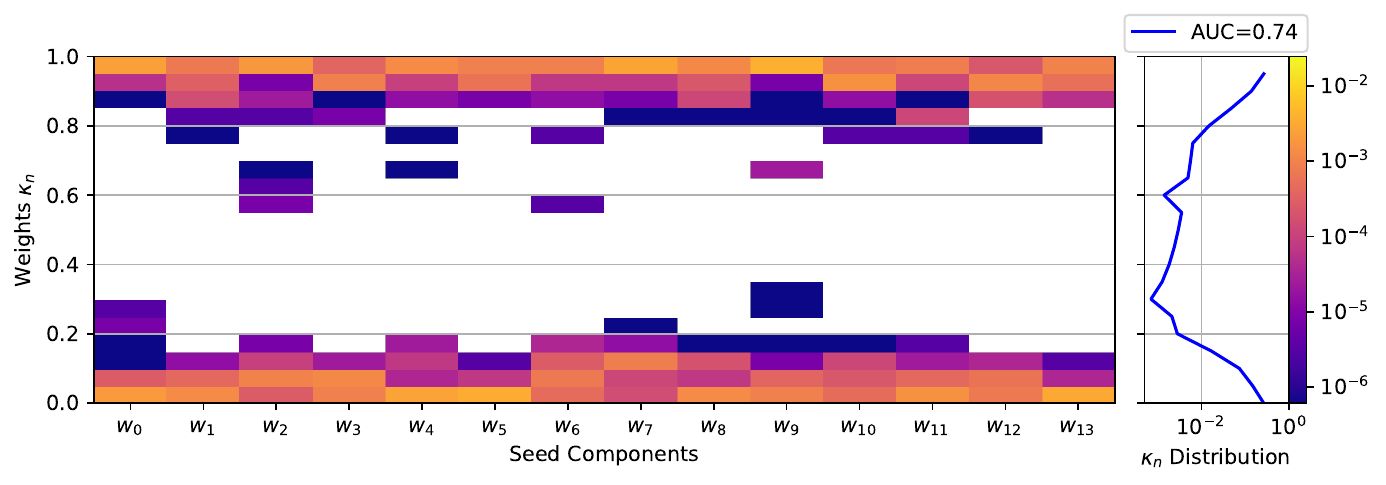}
        \vspace{-2em}
        \caption{ImageNet}
    \end{subfigure}%
    \vspace{-1em}
    \caption{Weights $\gamma_n$ of seed components}
    \label{fig:heat}
\end{figure}

In \autoref{fig:heat}, we show distribution of genome component values, aggregated across all candidates for each dataset. This aggregation is done for the experiments using $\{ \omega_{dcb}, \omega_{d2}\}$ The figures show a histogram for each component $w_n$, with the frequency in each bin being color coded on a log scale. The empty regions are zeros, i.e., no contribution. Additionally, we aggregate the usage across all seed components, giving us the $\kappa_n$-Distribution. With this distribution the usage can be shown more effectively, where the area under the curve (AUC) acts as a proxy for latent manipulation complexity.

From \autoref{fig:heat}, we observe a clear trend towards extreme values, with component values increasingly concentrated around smaller differences. Additionally, as dataset complexity increases, the spread of these concentrations widens, which is shown also in the change of AUC. However, this trend does not hold for CIFAR-10, indicating that other factors beyond data complexity affect the manipulation.

\begin{figure}[htbp]
    \centering
    \begin{subfigure}[b]{0.49\textwidth}
        \centering
        \includegraphics[width=\textwidth]{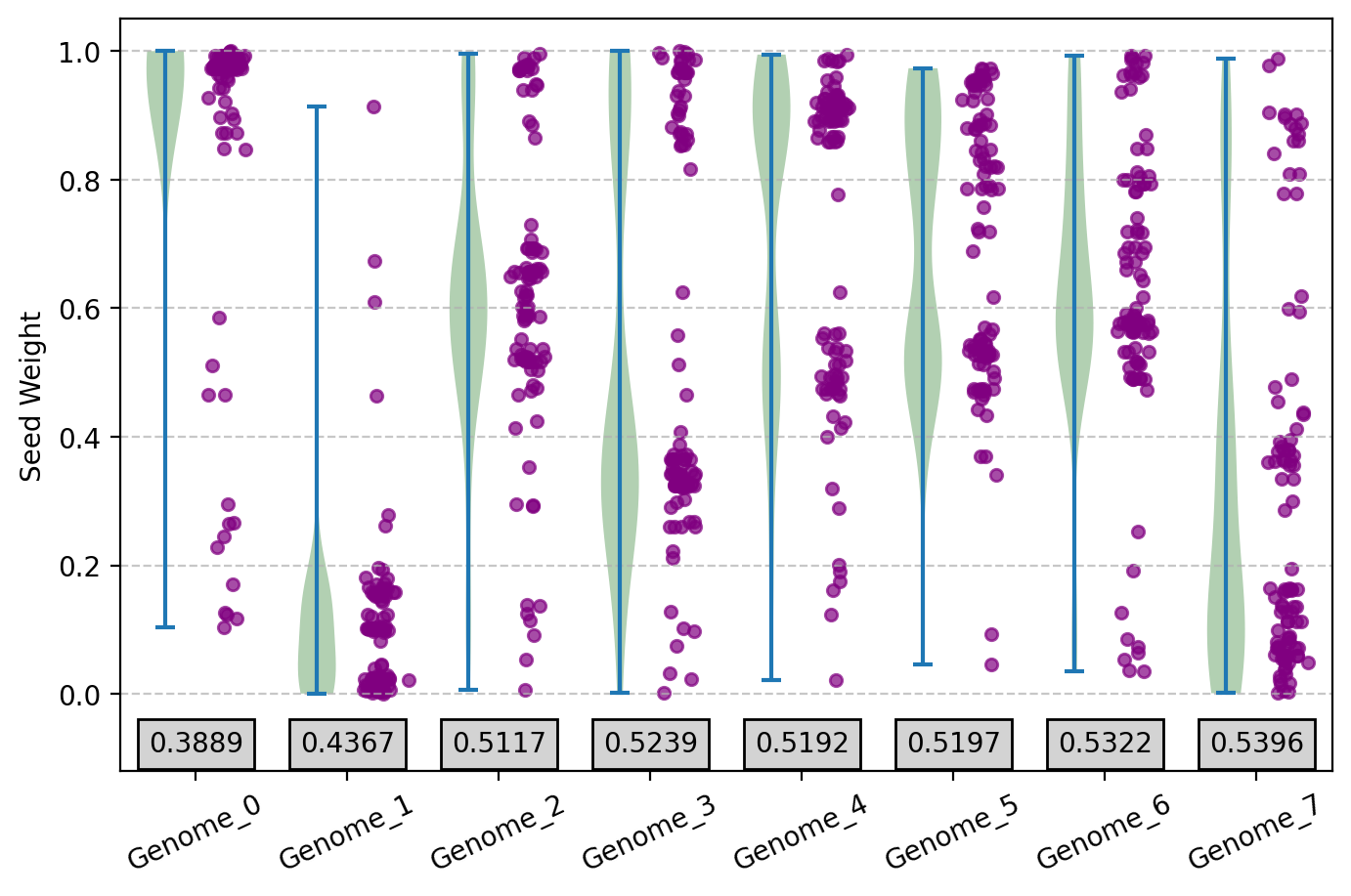}
        \vspace{-2em}
        \caption{$\omega = \{\omega_{dcb}\}$}
        \label{fig:heat_cif_a}
    \end{subfigure}%
    \begin{subfigure}[b]{0.49\textwidth}
        \centering
        \includegraphics[width=\textwidth]{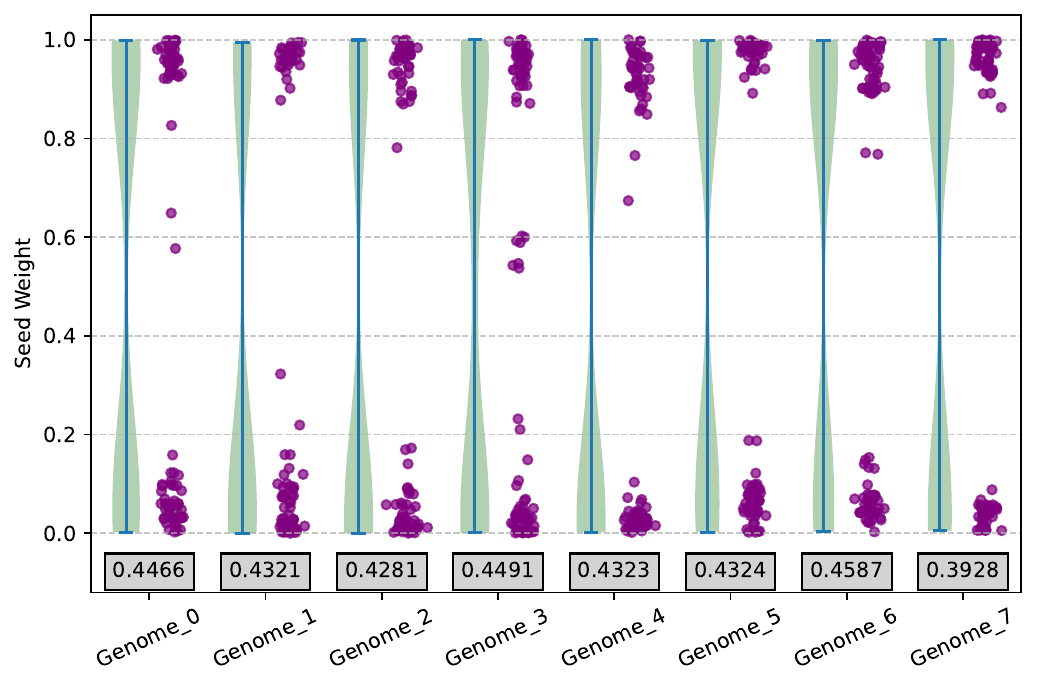}
        \vspace{-2em}
        \caption{$\omega = \{\omega_{dcb}, \omega_{d2}\}$}
        \label{fig:heat_cif_b}
    \end{subfigure}%
    \vspace{-1em}
    \caption{Genome usage in CIFAR-10, without and with diversity constraints.}
    \label{fig:heat_cif}
\end{figure}

To investigate how the usage of the genome changes, we remove the genome diversity objective. As expected, changing the objectives leads to changes in the resulting latent space usage (see \autoref{fig:heat_cif}). The plots in \autoref{fig:heat_cif} show the distribution of weights for each genome component in the CIFAR-10 case, as a violin and scatter plot. The number below each plot quantifies the uniformity of the distribution, with a value of 1 indicating perfect uniformity and 0 meaning all weights are identical.

Looking at the baseline with all objectives in \autoref{fig:heat_cif_b}, we observe a clear trend toward extreme values in the seed weights, as confirmed by \autoref{fig:heat_d}. Interestingly, some genome components (such as the ``coarser'' layers $w_1$ and $w_2$) show a preference for lower weight values.

When comparing this to \autoref{fig:heat_cif_a}, where the constraint on the genomes is removed, the distribution of weights changes noticeably. The uniformity measure is higher in this case, indicating a more distributed layer usage. In contrast to the baseline, we now see distinct preferences in some genome components, resulting in less divergent distributions.

\begin{tcolorbox}[boxrule=0pt,sharp corners,boxsep=2pt,left=2pt,right=2pt,top=2.5pt,bottom=2pt]
\begin{center}
\begin{minipage}[t]{0.99\linewidth}
\textbf{RQ\textsubscript{4} (latent space usage)}: \textit{\tool can use the latent space of its generator with great flexibility. Concerning datasets, the latent usage is more fine-grained as the dataset complexity increases, with an exception of CIFAR-10. Additionally, the choice of optimization objectives has a significant impact on how the latent space is used. In our case the latent manipulations resemble classical style-mixing operations with a novelty constraint used, whereas without diversity constraints, the manipulations have a more uniform distribution in interpolation weights.
}
\end{minipage}
\end{center}
\end{tcolorbox}

\subsection{Qualitative Analysis}\label{sec:qual}

While metrics and quantitative analysis allows for comparability of results, they often do not communicate the full picture adequately, especially with respect to image realism~\cite{lambertenghi2024assessing}. To showcase the performance differences between \tool and the baseline methods, we perform a qualitative analysis of the produced outputs by manually analyzing a few meaningful examples.

\begin{figure}[h]
    \centering
     \begin{subfigure}{0.32\textwidth}
        \centering
        \includegraphics[width=\linewidth]{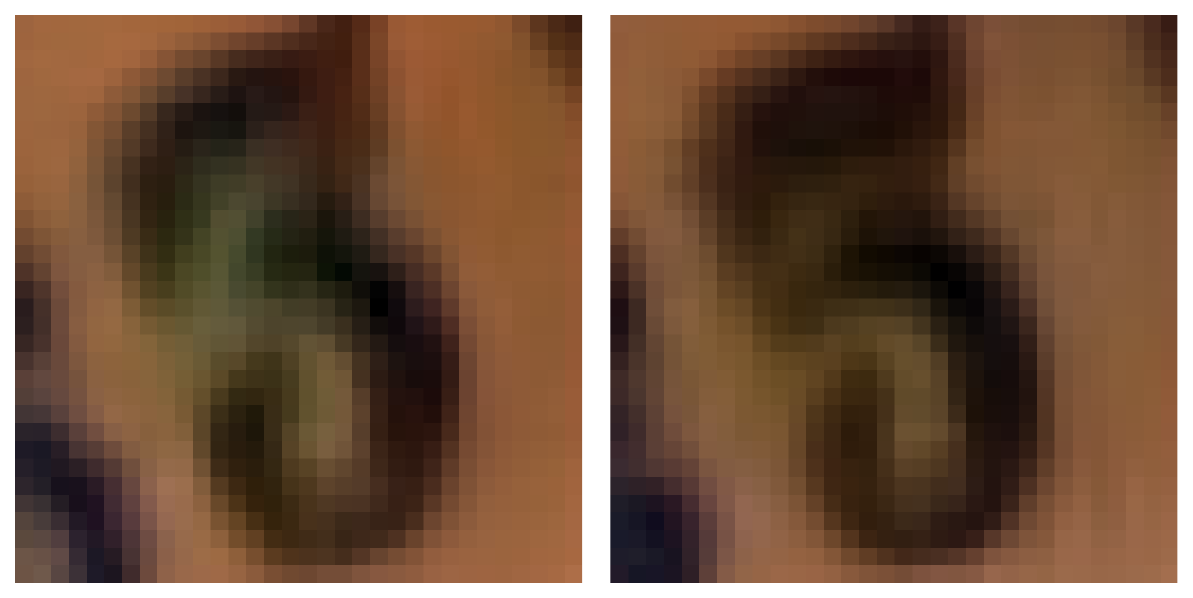}
        \includegraphics[width=\linewidth]{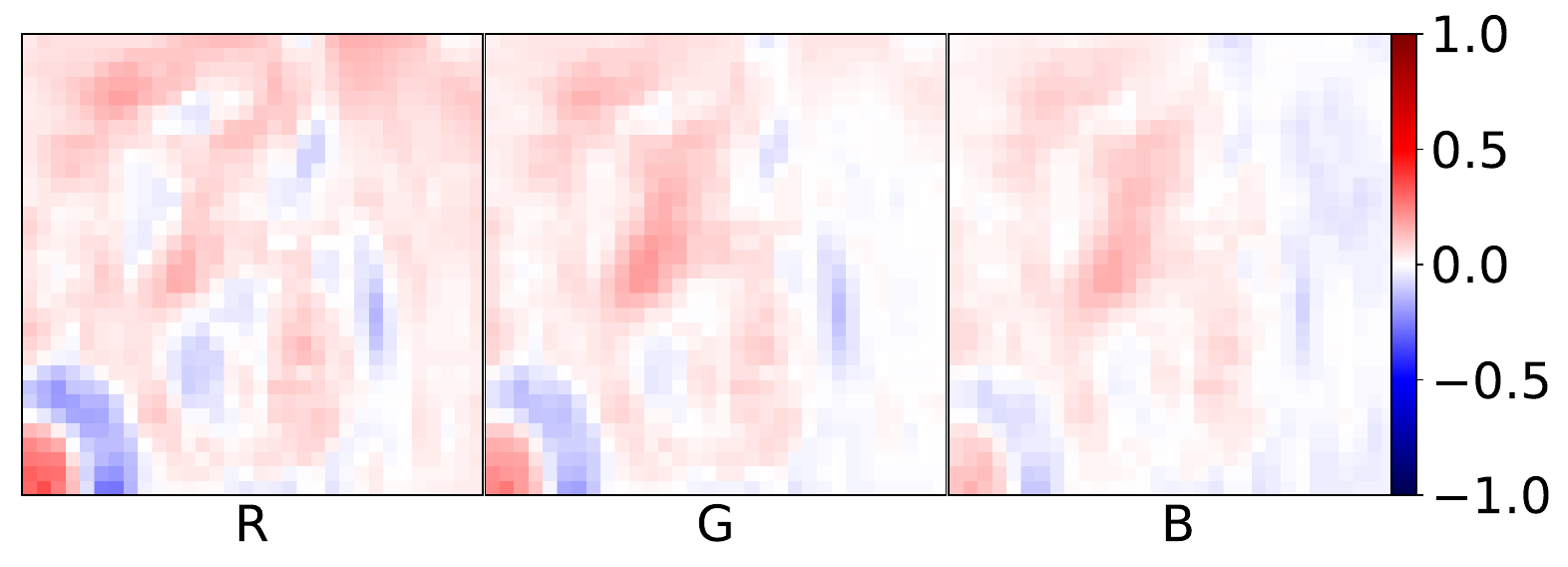}
        \caption{\textit{3} $\rightsquigarrow$ \textit{5} by \tool.}\label{fig:qual_svhn_a}
    \end{subfigure}
    \hfill
    \begin{subfigure}{0.32\textwidth}
        \centering
        \includegraphics[width=\linewidth]{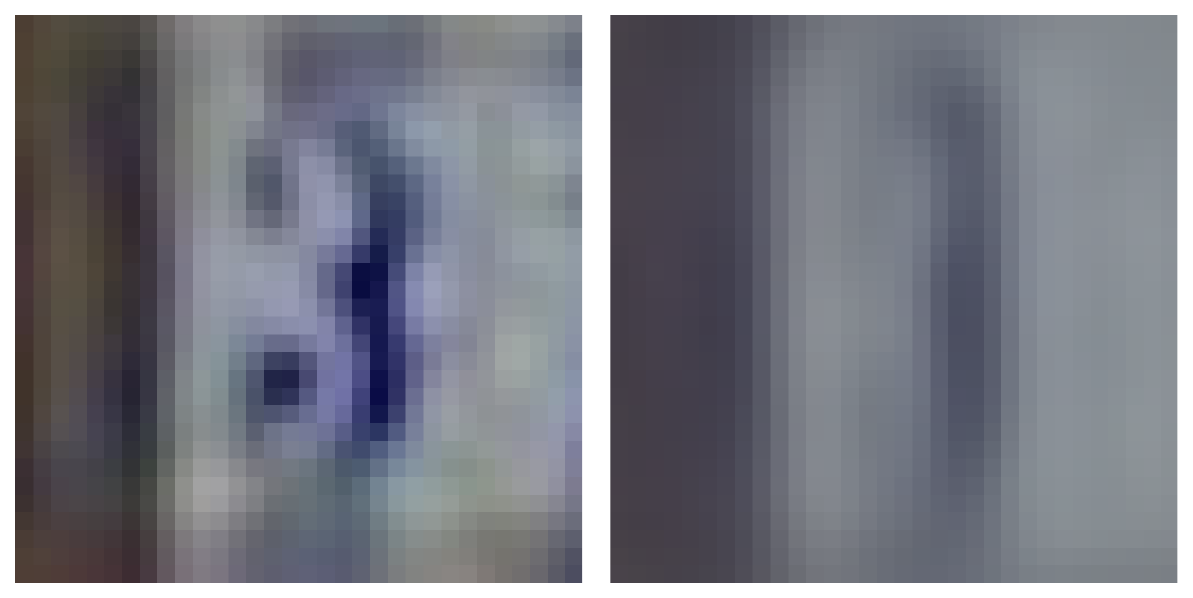}
        \includegraphics[width=\linewidth]{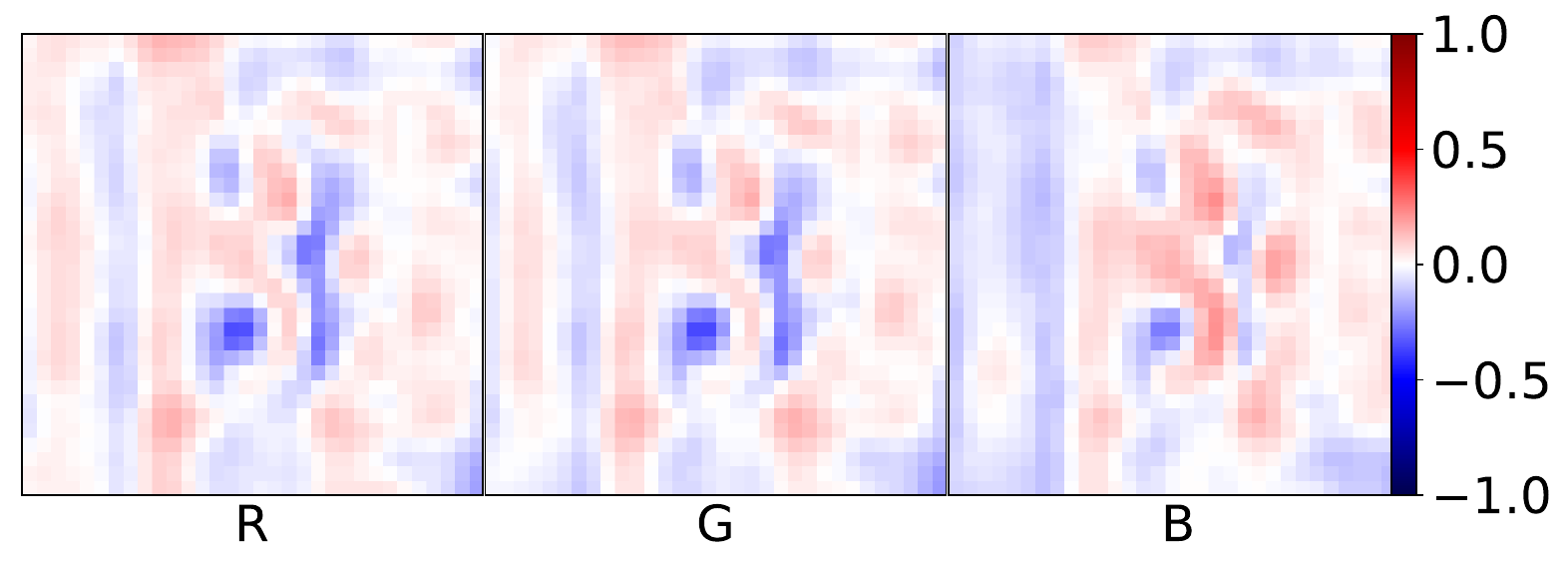}
        \caption{\textit{3} $\rightsquigarrow$ \textit{6} by Sinvad.}\label{fig:qual_svhn_b}
    \end{subfigure}
    \hfill
    \begin{subfigure}{0.32\textwidth}
        \centering
        \includegraphics[width=\linewidth]{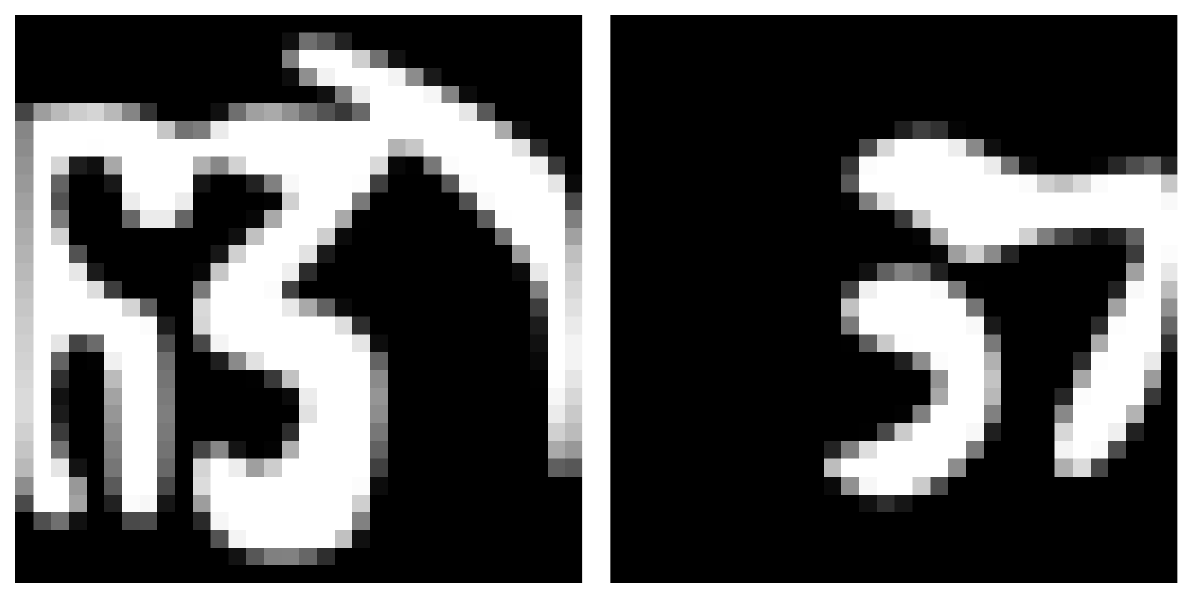}
        \includegraphics[width=\linewidth]{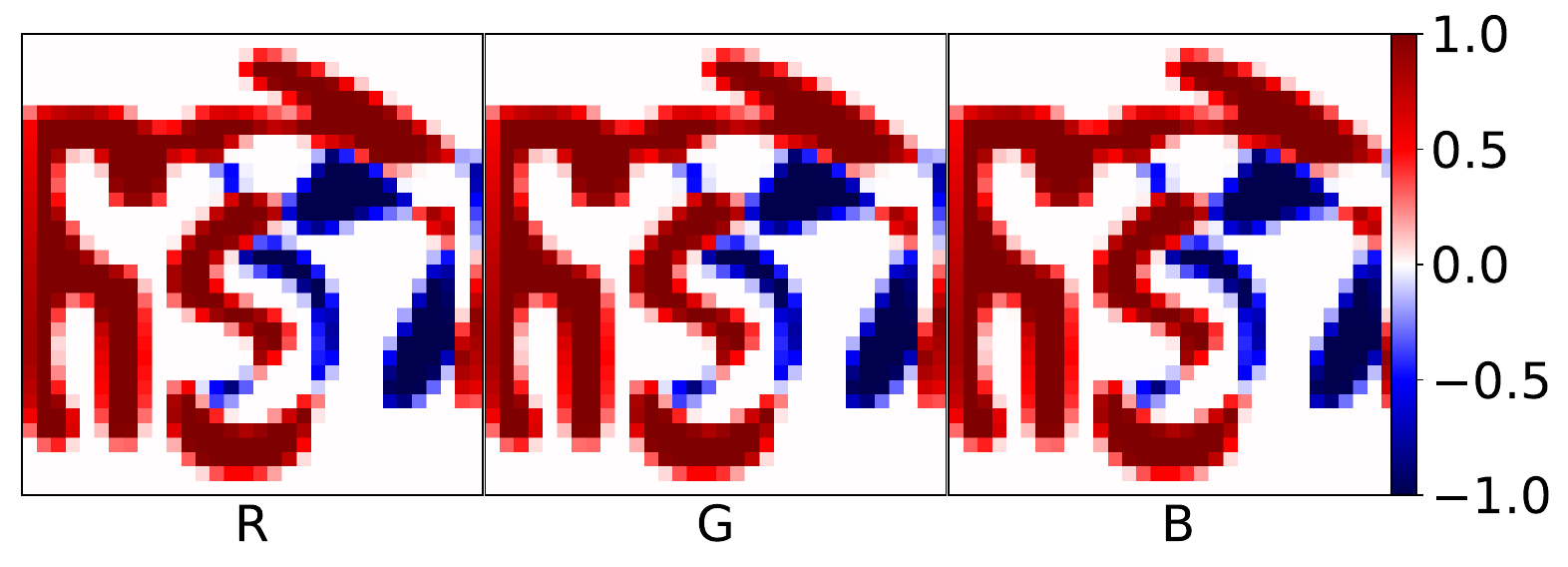}
        \caption{\textit{3} $\rightsquigarrow$ \textit{1} by DeepJanus.}\label{fig:qual_svhn_c}
    \end{subfigure}
    \vspace{0.5em}
    \begin{subfigure}{0.49\textwidth}
        \centering
        \includegraphics[width=\linewidth]{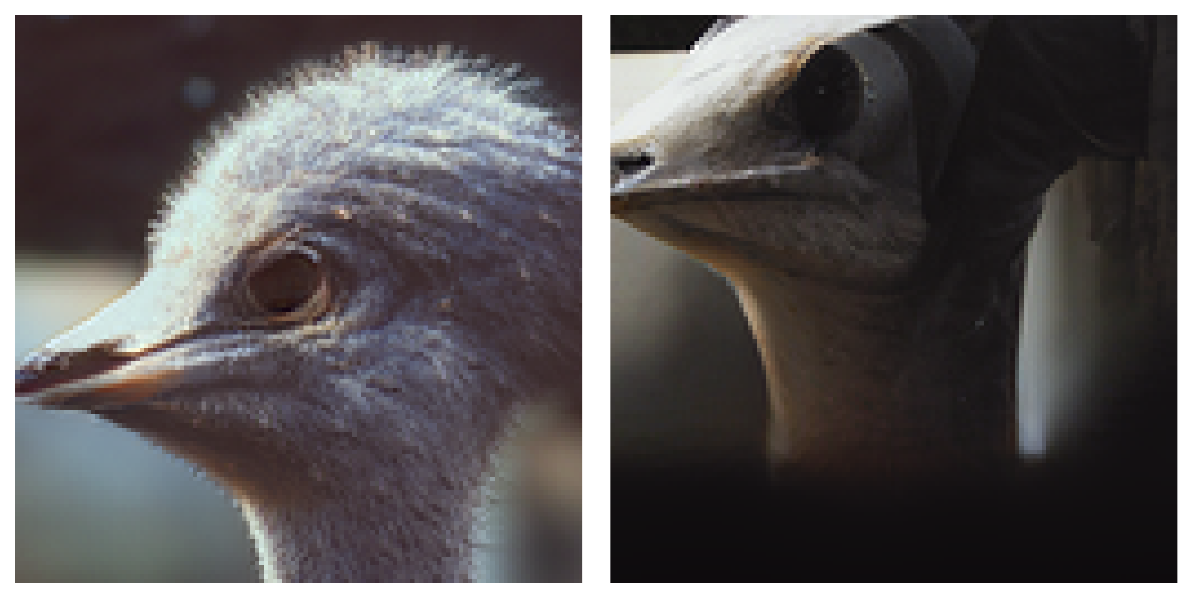}
        \includegraphics[width=\linewidth]{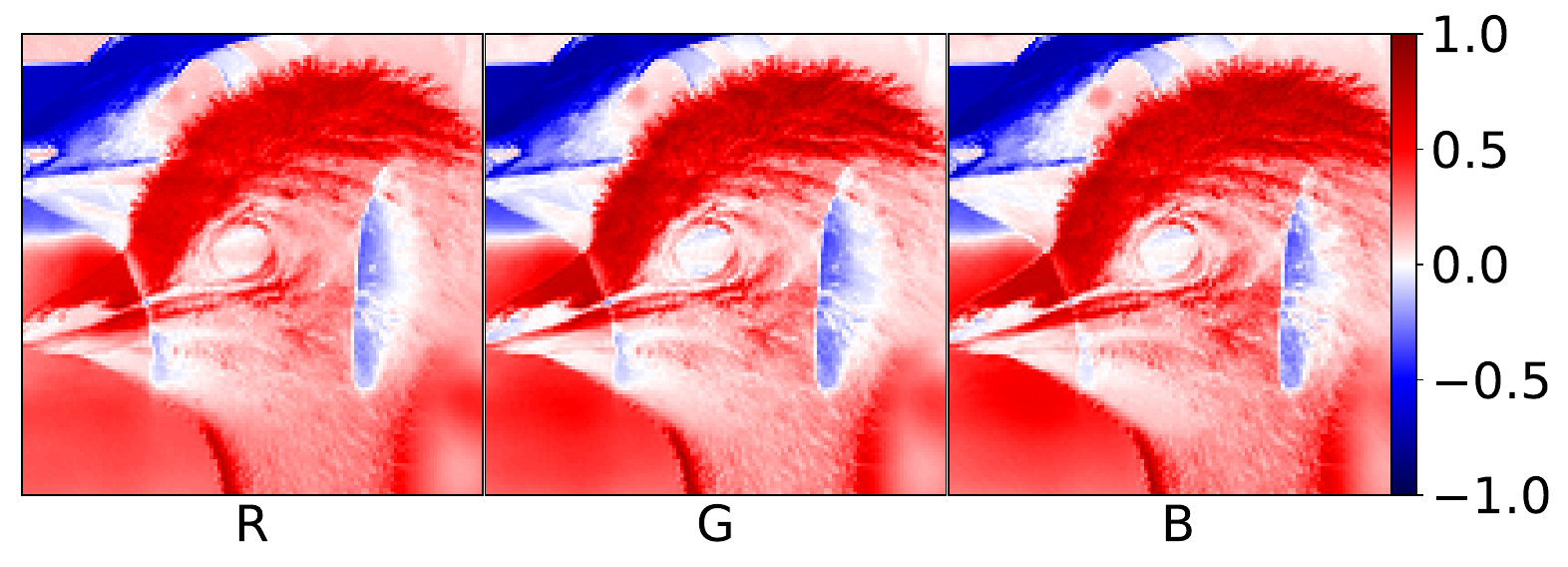}
        \caption{\textit{ostrich} $\rightsquigarrow$ \textit{electric ray} by \tool.}\label{fig:qual_im_a}
    \end{subfigure}
    \hfill
    \begin{subfigure}{0.49\textwidth}
        \centering
        \includegraphics[width=\linewidth]{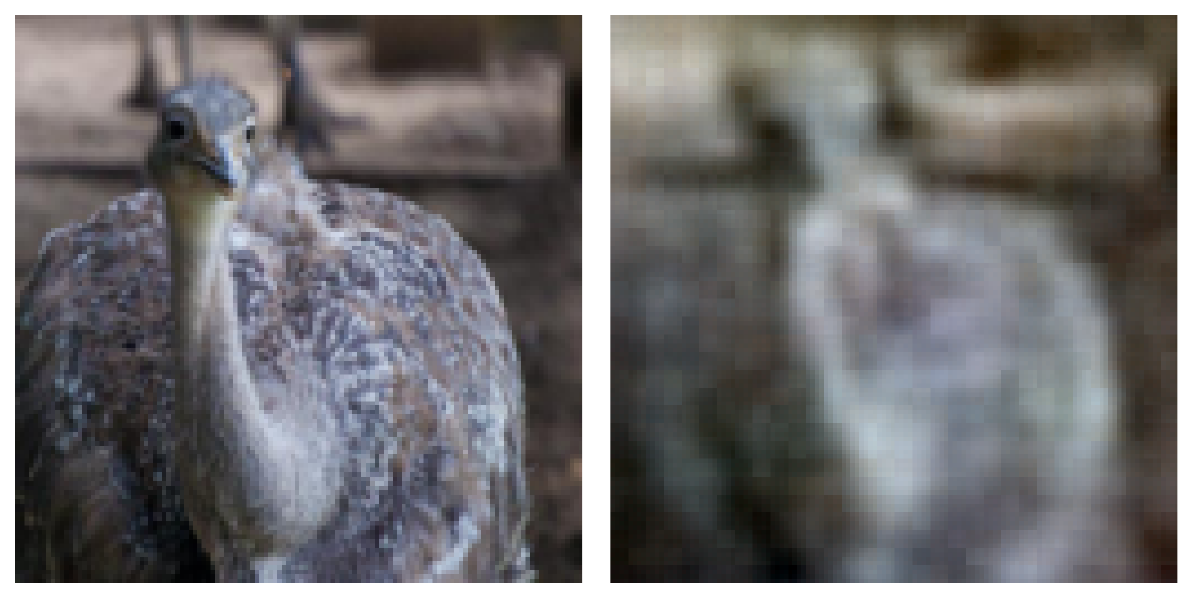}
        \includegraphics[width=\linewidth]{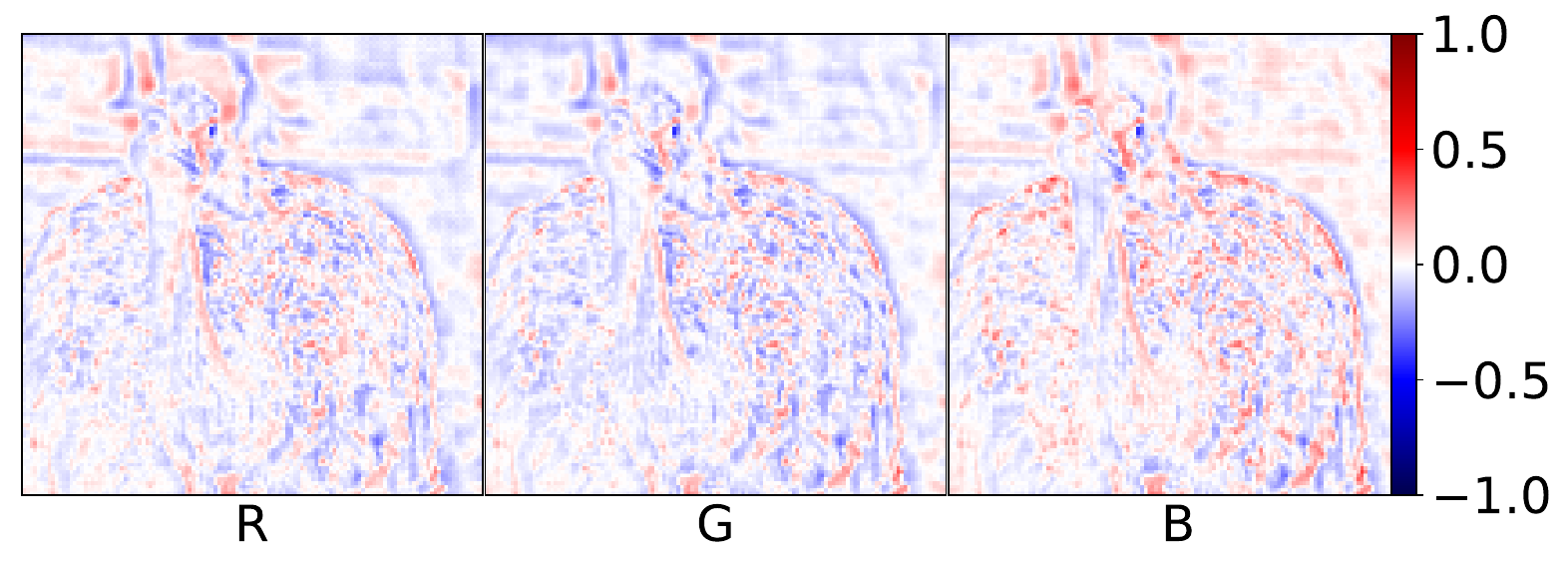}
        \caption{\textit{ostrich} $\rightsquigarrow$ \textit{stringray} by Sinvad.}\label{fig:qual_im_b}
    \end{subfigure}
    \caption{\textbf{SVHN \& ImageNet:} Initial to Final Candidate Comparison with Channel-wise Differences.}
    \label{fig:qual}
\end{figure}

The first interesting aspect relates to the Laplacian Variance of image differences seen in \autoref{eq:lap_var}. This numerical measure it is not widely used and therefore needs more explanation and positioning. Especially in comparing \tool to Sinvad it is useful, while it fails for the comparison with DeepJanus. As described earlier with this measure the type of change in the image across the optimization process is quantified. Low Laplacian variance in the image differences here means that the image gets gradually blurred, not producing functional differences in the output candidate (\autoref{fig:qual_svhn_b} \& \autoref{fig:qual_im_b}). This can easily be seen in the examples for Sinvad, where with more complex data, this phenomenon gets more prominent. In contrast, \tool does not blur the original seed images, rather produces functionally different outputs, even if those outputs may no longer convey clear class information to the human observer. 

\subsection{Threats to Validity}\label{sec:ttv}
\subsubsection{Internal validity}
All experiments of \tool and the baselines have been conducted on the same computational budget, based on SUT predictions.
Additionally, regarding the StyleGAN models, we utilized pre-trained models available from the literature~\cite{sauer2022styleganXL, karras2020analyzing}. When not available, we trained the StyleGAN models using the scripts available in the replication package of the original paper~\cite{karras2020analyzing}, as it is difficult to envision less threat-prone approaches. In the human evaluation study, we included attention questions to ensure that participants were engaging truthfully \cite{sorokin2008utility}. Since both DeepJanus and Sinvad require input images, rather than generating inputs like \tool, the input selection process was randomized. This approach aims to reduce bias and ensures a more fair comparison with \tool.

\subsubsection{External validity}
The limited number of ML systems included in our evaluation poses a threat to the generalizability of our results. We addressed this issue by incorporating a variety of datasets with increasing complexity and high-performing models from related literature.
We demonstrated the usefulness of the StyleGAN architecture~\cite{karras2020analyzing, sauer2022styleganXL}; however, other style-based architectures~\cite{karras2019style, karras2021alias} may also yield promising results.
\section{Discussion}\label{sec:discussion}

\head{Latent feature mixing enables semantic control during test generation}
\tool proved effective for boundary identification in all considered benchmarks and SUTs. The effectiveness can be attributed to the disentangled latent space representation, which produces high quality boundary inputs if explored effectively. On the contrary, entangled latent space representations often result in blurred effects applied to the original images, with no benefits in functional coverage.
Interestingly, the boundary case quality seems to be independent of the SUT, where more ambiguity in the predicted class probabilities does not result in worse boundary candidates, as it is observable with the baseline methods (see \autoref{fig:comp}). 
However, our results show that both the SUT quality and benchmark complexity affect the manipulations in \tool's generator, as in the case of CIFAR-10 and ImageNet. Our experiments also show that imposing diversity constraints in the objective function affect the latent space manipulation, which in turn affects the final quality and validity of the generated boundary candidates.
Concerning the relevance of generated boundary inputs, our results show that \tool outperforms the baselines in all datasets as it produces inputs that represent the target class constraint (i.e., they do not ``escape'' the given boundary). Overall, \tool maintains a competitive scores across datasets, while staying within the distribution of the data domain under test. In contrast, Sinvad and DeepJanus maintains reasonable effectiveness in smaller datasets such as MNIST and FashionMNIST, but deteriorate with more complex data, where they either produce corrupted or out-of-distribution data. 

\head{Targeted boundary exploration improves functional coverage}
Concerning coverage, it is important to consider that for some classes only a subset of all possible labels is meaningful, whereas high coverage would suggest poor control over test case generation. As an example, the digit 7 in MNIST has meaningful boundary candidates in the numbers 1 or 5, whereas an 8 probably is not bounding to the decision region of 7s. In this case, \tool's targeted exploration overcomes existing tools, and it proves to outperform the competitors especially when dataset complexity increases (\autoref{fig:lcer}), thanks to its targeted exploration. 

\head{Generating well-defined, unambiguous boundary inputs for high-resolution datasets remains an open challenge}
\tool produces inputs with high validity rates, provided the generated images depict a class that is perceptible to human observers. In terms of boundary preservation, \tool generally performs well except in the case of ImageNet. Unlike Sinvad, which mainly blurs existing dataset images, \tool generates functionally novel inputs that may sometimes appear ambiguous to humans. For ImageNet, the relatively high resolution ($128 \times 128$) allows the blur introduced by Sinvad to retain enough visual cues for human evaluators to identify the original class. In contrast, \tool may generate objects that are less recognizable or entirely unfamiliar, given the feature mixing between two classes. This highlights an open challenge: as dataset complexity increases, assessing the validity of generated test cases becomes more difficult for human evaluators, especially for classes with low to no semantic affinity.
In such contexts, alternative evaluation methods such as using large language models as judges~\cite{gu2025surveyllmasajudge} might be of interest to corroborate the human assessment.
\section{Related Work}\label{sec:related}
While testing objectives can vary drastically, the methodologies used to achieve the testing objective can be grouped in three families. These families of ML test generation methodologies are model-based input manipulation, raw input manipulation, and latent space manipulation. We overview the main propositions next to clarify the positioning of \tool.

\subsection{Model-based Input Manipulation}\label{mim}
Model Input Manipulation (MIM) techniques leverage a model of the input domain to generate test inputs, similar to conventional model-driven engineering practices that uphold compliance with domain-specific constraints~\cite{Abdessalem-ASE18-1,Abdessalem-ASE18-2,Abdessalem-ICSE18,Gambi:2019:ATS:3293882.3330566, riccio2020deepjanus,isa}.

The manipulation occurs on the model, which is subsequently reconverted to the original format~\cite{larman1998applying}.
MIM techniques operate within a restricted input space, specifically the control parameters of the model representation. These techniques enhance the realism of the produced outputs by implementing appropriate model constraints.

Several search-based MIM approaches have been applied to DL-based image classifiers. DeepHyperion~\cite{zohdinasab2021deephyperion} uses the MAP-Elites Illumination Search algorithm~\cite{DBLP:journals/corr/MouretC15} to explore the feature space of the input domain and identify misbehavior-inducing features. DeepMetis~\cite{2021-Riccio-ASE} a MIM approach that generates inputs that behave correctly on original DL models and misbehave on mutants obtained through injection of realistic faults~\cite{2020-Humbatova-ICSE}, which can be useful to enhance the mutation killing ability of a test set.
DeepJanus~\cite{riccio2020deepjanus} is the MIM approach most related to this work since it performs boundary testing of DL systems. Therefore, we performed an explicit empirical comparison with the DeepJanus approach in this work.

However, a significant limitation of MIM approaches is their reliance on the availability of a high-quality model representation for the specific input domain, which is manually crafted~\cite{2023-Riccio-ICSE}.
Unlike MIM techniques, \tool leverages a generative network to learn the distribution of the input domain. This approach is largely automated and requires no labeling or other cost, except for hyperparameter tuning. This characteristics of \tool broadens its applicability across various domains.

\subsection{Raw Input Manipulation}\label{rim}

Raw input manipulation (RIM) techniques involve modifying an image's original pixel space to create a new input by perturbing the pixel values. 
RIM techniques aim to produce minimal, often imperceptible changes to original to trigger misbehavior in the DL system \cite{liu2022deepboundary,zhang2020deepsearch, zohdinasab2021deephyperion, kurakin2018adversarial, croce2020minimally}. These methods do not focus on boundary analysis and target different aspects of testing, such as data augmentation or adversarial attacks, which are not directly aligned with our goal. Our method is a \textit{functional} test generator, differing from adversarial testing in both goals and techniques. Functional testing creates new, valid, in-distribution inputs to evaluate a DNN's generalization. In contrast, adversarial testing adds minor perturbations to original inputs to test \textit{robustness} \cite{croce2020minimally}. Given these distinct objectives and methods, direct comparisons are inappropriate. However, for completeness, we describe the main propositions next.

DeepXplore~\cite{pei2017deepxplore} employs various techniques, including occlusion, light manipulation, and blackout to cause misbehavior. These perturbations are intended to improve neuron coverage within the DL system.
DLFuzz~\cite{guo2018dlfuzz} introduces noise to the seed image to increase the likelihood of system misbehavior. DLFuzz generates adversarial inputs for DL systems without relying on cross-referencing other similar DL systems or manual labeling. 
DeepTest~\cite{deeptest} alters the images using synthetic affine transformation from the computer vision domain, such as blurring and brightness adjustments, to create simulated rain/fog effects.

RIM techniques are limited to modifying existing inputs and they cannot thoroughly explore the input domain and its boundaries, while generative DL models can sample novel inputs from the data distribution.
Moreover, the manipulated images might not always represent real-world functional inputs, e.g., images with artificial artifacts at the corners or unnatural lighting conditions generated by DeepXplore.
Consequently, such techniques are more suitable for security and robustness testing rather than for functional testing~\cite{2023-Riccio-ICSE}.

Differently, our technique targets functional testing, specifically boundary value analysis of ML systems. We achieve this by manipulating the latent space of a StyleGAN to efficiently find test cases that expose behavioral changes in the SUT.

\subsection{Latent Space Manipulation}\label{lsm}

Latent space manipulation techniques generate new inputs by learning and reconstructing the underlying distribution of the input data. The most commonly used techniques are Variational Autoencoders (VAE)~\cite{kingma2013auto} and Generative Adversarial Networks (GAN)~\cite{goodfellow2020generative}.

Sinvad~\cite{kang2020sinvad, sinvad-tosem} constructs the input space using VAE and navigates the latent space by adding a random value sampled from a normal distribution to a single element of the latent vector. Sinvad aims to explore the latent space by maximizing either the probability of misbehaviors, estimated from the softmax layer output, or by surprise coverage~\cite{kim2019guiding}. 

The Feature Perturbations technique~\cite{dunn2021exposing, DBLP:journals/corr/abs-2001-11055} involves injecting perturbations into the output of the generative model’s first layers, which represent high-level features of images. These perturbations can affect various characteristics of the image, such as shape, location, texture, or color. 
DeepRoad~\cite{deeproad} generates driving images using Generative Adversarial Networks (GANs) for image-to-image translation.

CIT4DNN~\cite{dola2024cit4dnn} combines VAE and combinatorial testing~\cite{cit605761}. This allows the systematic exploration and generation of diverse and infrequent input datasets.
CIT4DNN partitions latent spaces to create test sets that contain a wide range of feature combinations and rare occurrences. 
A recently proposed technique, Instance Space Analysis, aims to pinpoint the critical features of test scenarios that impact the detection of unsafe behavior~\cite{isa}.

Unlike conventional latent space manipulation techniques, our approach leverages the richer and more complex latent space of StyleGAN for boundary testing of ML systems. While existing state-of-the-art methods are often constrained by limited data complexity, our framework addresses this limitation by incorporating more complex datasets, facilitating better transferability to real-world scenarios. Furthermore, we integrate feedback from the SUT in the form of model predictions to guide manipulations toward more promising regions of the decision space. We introduce \tool, a tool for ML testing, and demonstrate its effectiveness through a boundary testing case study.
\section{Conclusion \& Future Work}

In this work, we present \tool, a tool for targeted boundary testing of ML classifiers by identifying inputs near decision boundaries. Our empirical analysis demonstrates that \tool outperforms existing methods such as DeepJanus~\cite{riccio2020deepjanus} and Sinvad~\cite{kang2020sinvad}, particularly in complex data domains, by leveraging latent space manipulations and incorporating SUT behavior into the search process. Unlike DeepJanus, which relies on model representations of inputs, and Sinvad, which suffers from limited control over generative manipulations, \tool effectively balances control, fidelity, and performance in generating meaningful boundary test cases.

Future work will investigate the interplay between classifier quality and latent space complexity. Especially the concept of the boundary to a validity domain is seldom talked about in related literature, as it has an especially hard oracle problem. Another direction for future work is the change of manipulator technologies to other generators such as diffusion- or transformer-based generators. Additionally, unlike  previous methods, using \tool more complex datasets can be considered, making future work increasingly more relevant to real-world problems.

\section{Data Availability}\label{sec:da}
The experiment codebase, analysis scripts and all artifacts generated for this work can be found in the replication package~\cite{replication-package}. Artifacts that were generated by other works are linked accordingly. 

\section*{Acknowledgements}
\addcontentsline{toc}{section}{Acknowledgements}
This research was partially supported by the Bavarian Ministry of Economic Affairs, Regional Development and Energy, and the Practical Research Experience Program (PREP) of the Technical University of Munich. Vincenzo Riccio is supported by the Project SOP CUP N. H73C22000890001 PNRR M4 C2 I1.3 ``SEcurity and RIghts in the CyberSpace (SERICS)'' PE0000014 PE7  funded by Next-Generation EU.

\balance
\bibliographystyle{ACM-Reference-Format}
\bibliography{paper}
\appendix
\section{Study}
\begin{figure}[h]
    \centering
    \includegraphics[width=0.55\textwidth]{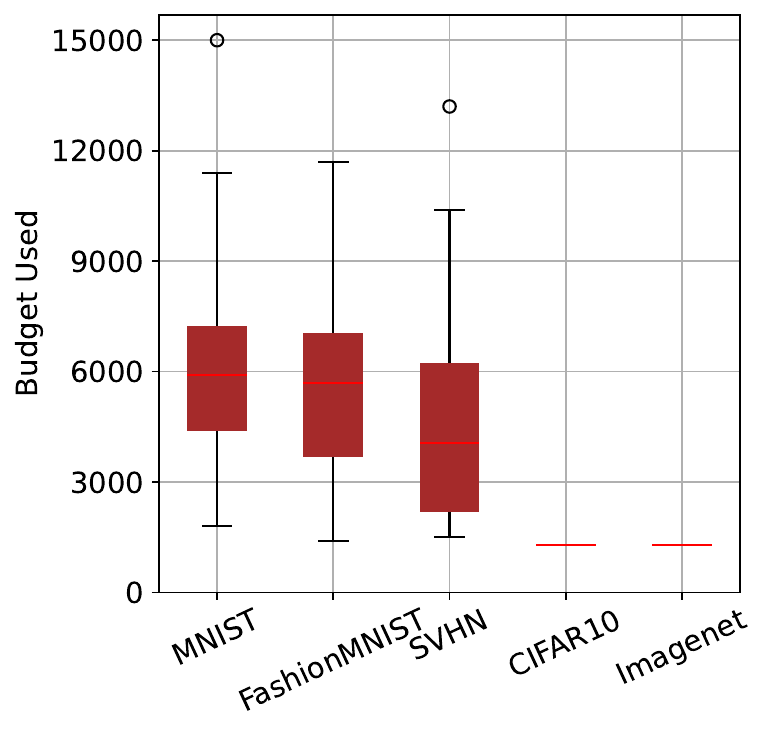}
    \caption{Sinvad Iterations Used}
    \label{fig:sinv_it}
\end{figure}

\begin{figure}
    \centering
    \begin{subfigure}[b]{0.19\textwidth}
    \centering
    \includegraphics[width=\linewidth]{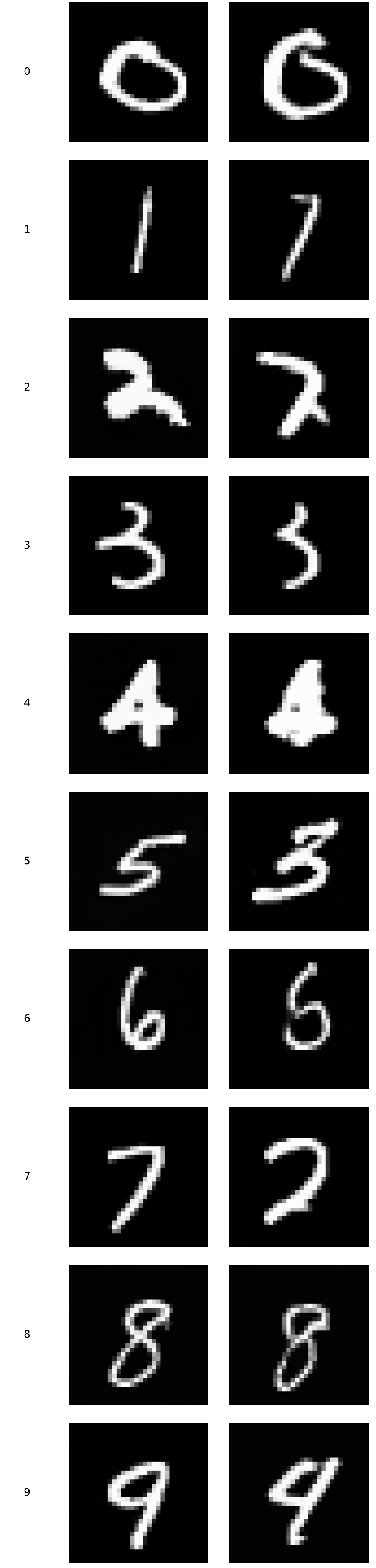}
    \caption{MNIST}
    \end{subfigure}
    \begin{subfigure}[b]{0.19\textwidth}
    \centering
    \includegraphics[width=\linewidth]{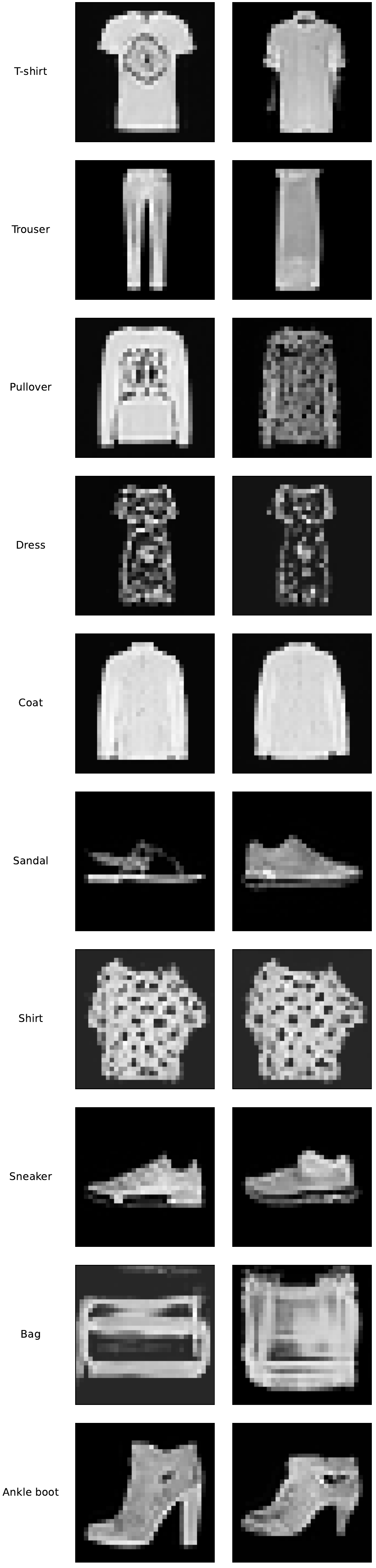}
    \caption{FashionMNIST}
    \end{subfigure}
    \begin{subfigure}[b]{0.19\textwidth}
    \centering
    \includegraphics[width=\linewidth]{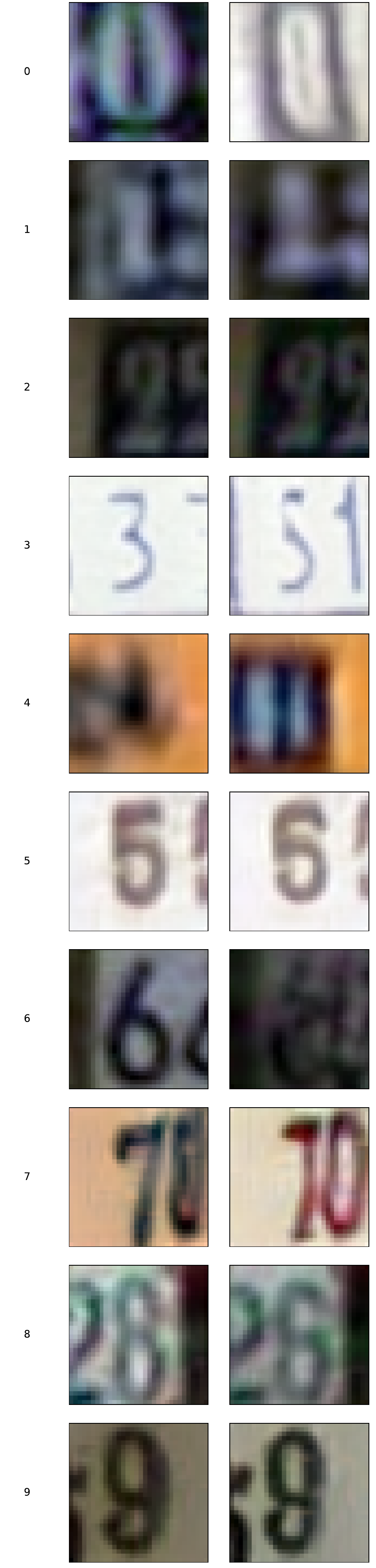}
    \caption{SVHN}
    \end{subfigure}
    \begin{subfigure}[b]{0.19\textwidth}
    \centering
    \includegraphics[width=\linewidth]{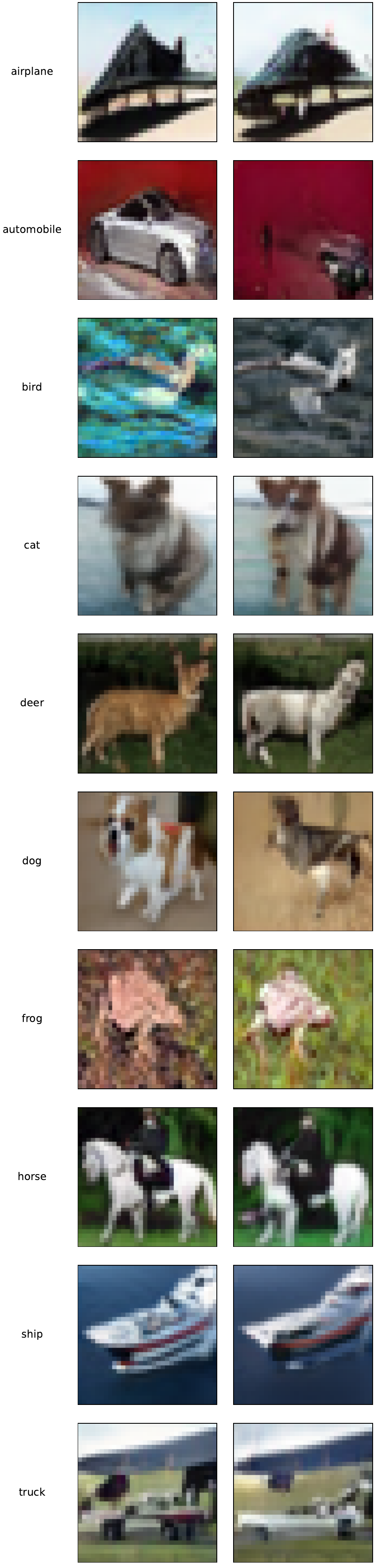}
    \caption{CIFAR-10}
    \end{subfigure}
    \begin{subfigure}[b]{0.19\textwidth}
    \centering
    \includegraphics[width=\linewidth]{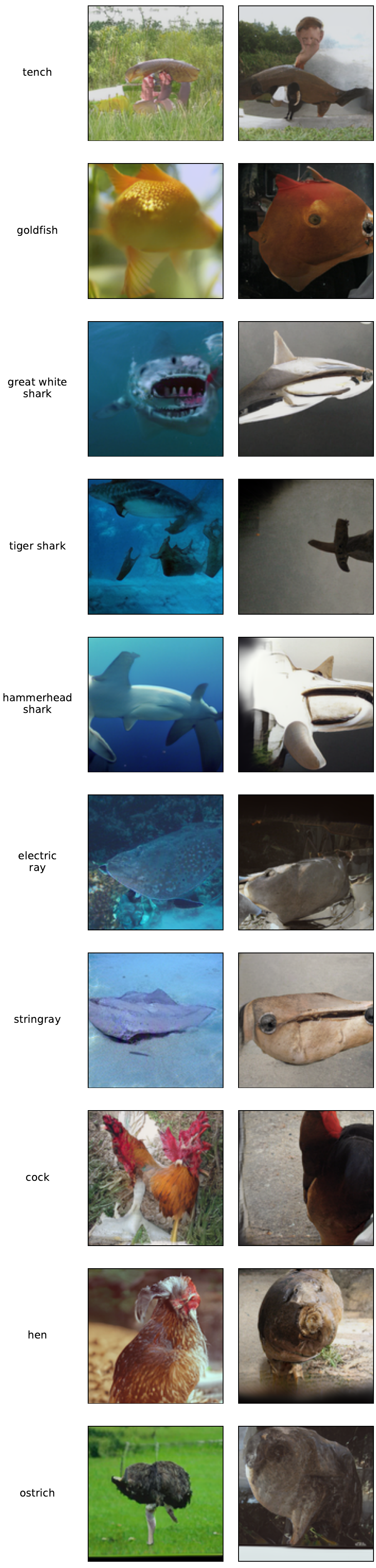}
    \caption{ImageNet}
    \end{subfigure}
    \caption{\tool original vs final candidate examples.}\label{fig:smoo_row_comp}
\end{figure}
\begin{figure}
    \centering
    \begin{subfigure}[b]{0.19\textwidth}
    \centering
    \includegraphics[width=\linewidth]{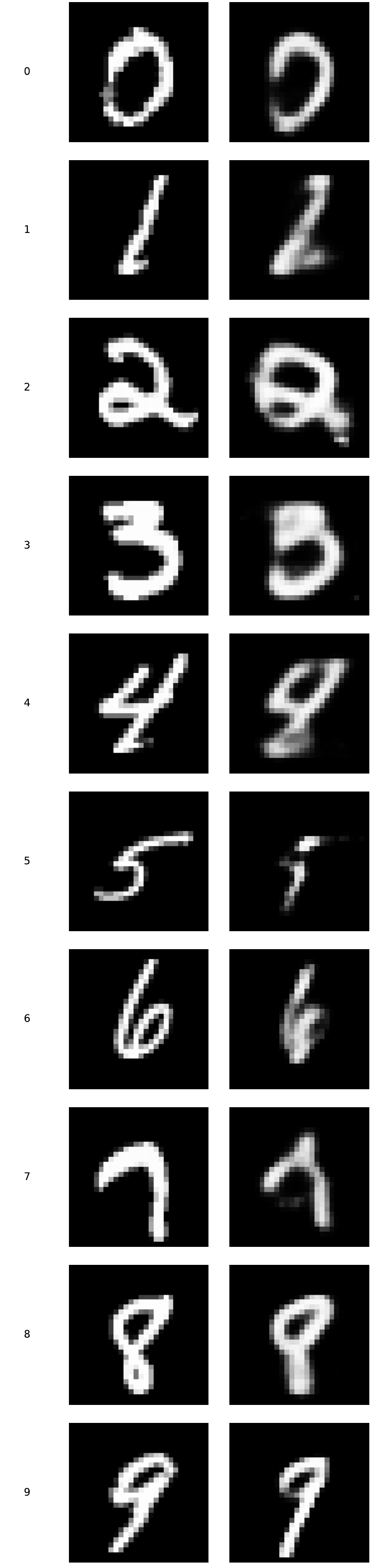}
    \caption{MNIST}
    \end{subfigure}
    \begin{subfigure}[b]{0.19\textwidth}
    \centering
    \includegraphics[width=\linewidth]{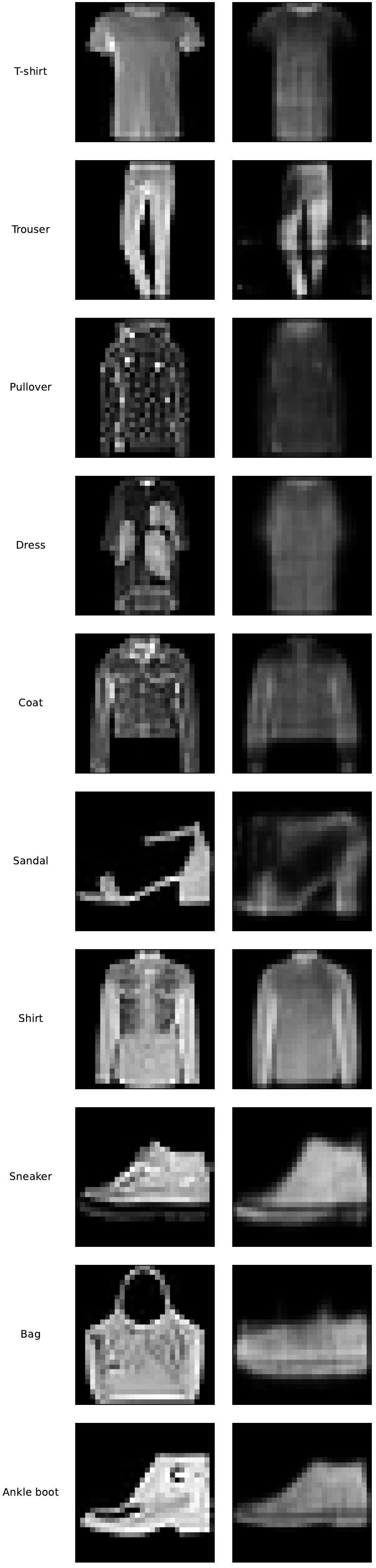}
    \caption{FashionMNIST}
    \end{subfigure}
    \begin{subfigure}[b]{0.19\textwidth}
    \centering
    \includegraphics[width=\linewidth]{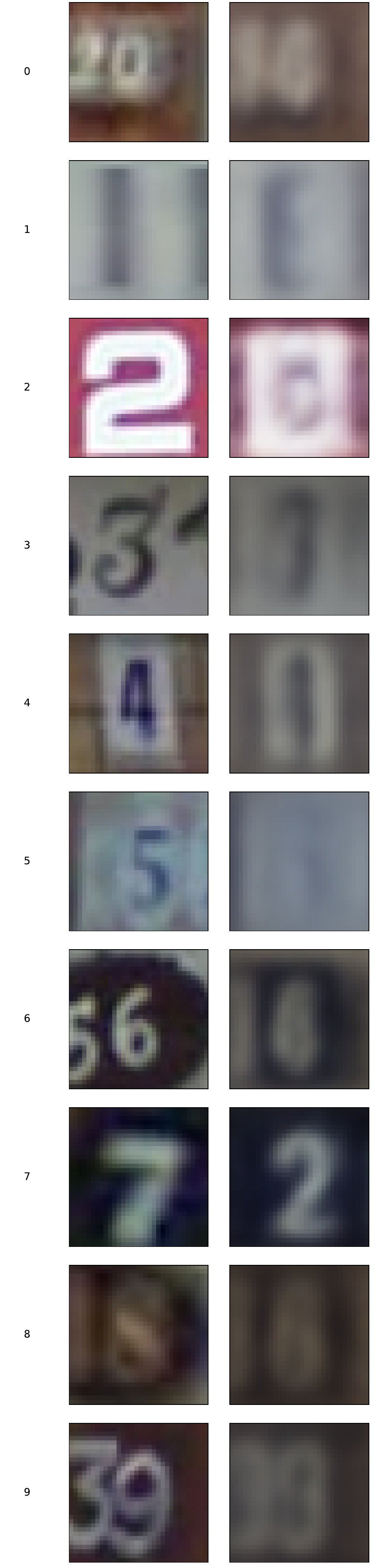}
    \caption{SVHN}
    \end{subfigure}
    \begin{subfigure}[b]{0.19\textwidth}
    \centering
    \includegraphics[width=\linewidth]{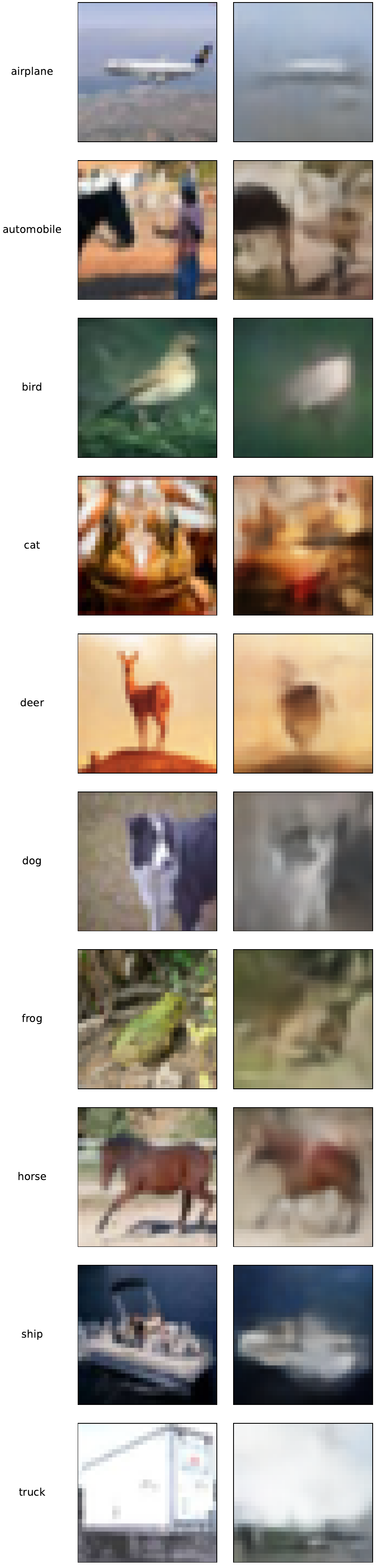}
    \caption{CIFAR-10}
    \end{subfigure}
    \begin{subfigure}[b]{0.19\textwidth}
    \centering
    \includegraphics[width=\linewidth]{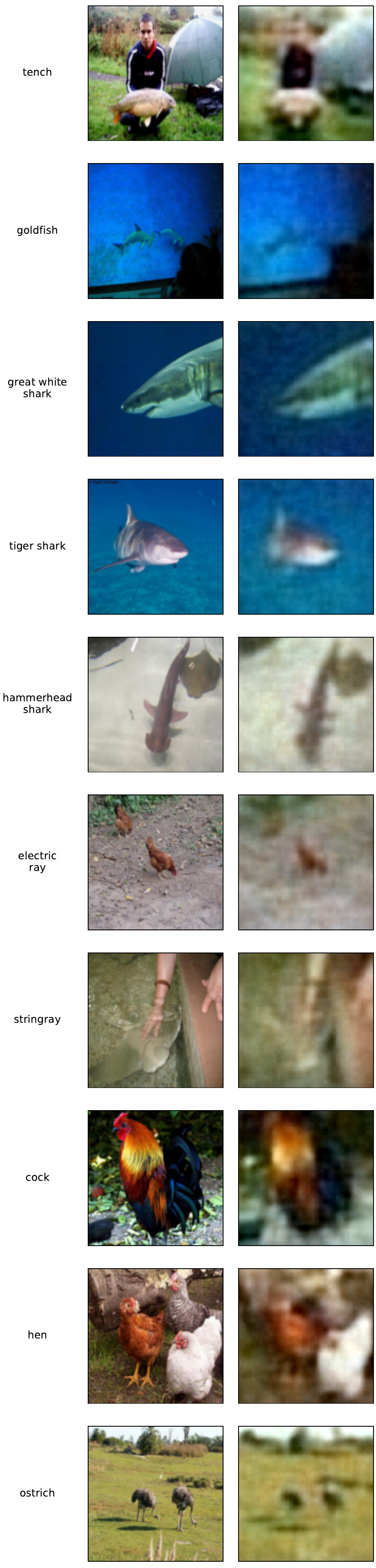}
    \caption{ImageNet}
    \end{subfigure}
    \caption{Sinvad original vs final candidate examples.}\label{fig:sinv_row_comp}
\end{figure}
\begin{figure}
    \centering
    \begin{subfigure}[b]{0.19\textwidth}
    \centering
    \includegraphics[width=\linewidth]{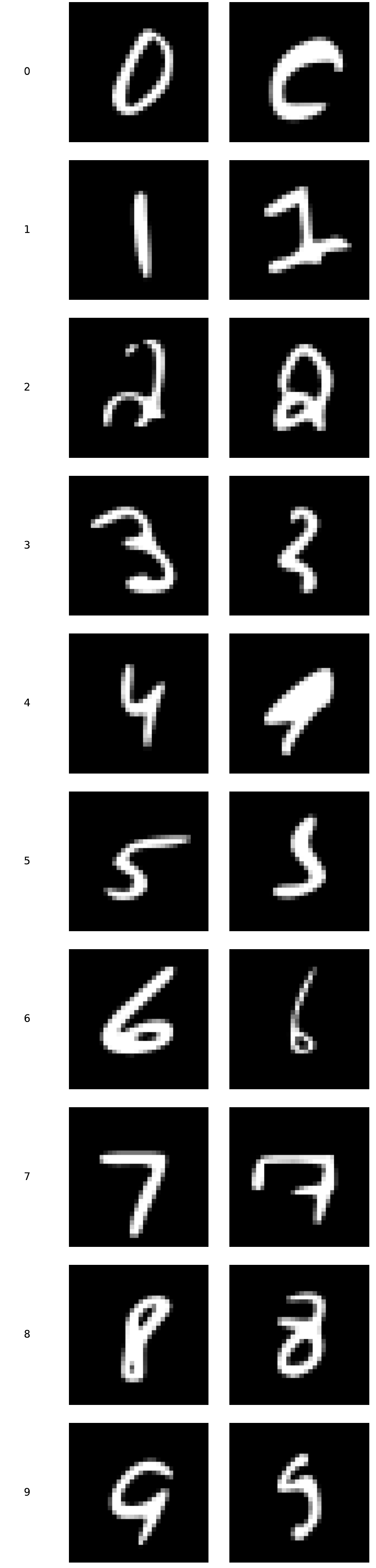}
    \caption{MNIST}
    \end{subfigure}
    \begin{subfigure}[b]{0.19\textwidth}
    \centering
    \includegraphics[width=\linewidth]{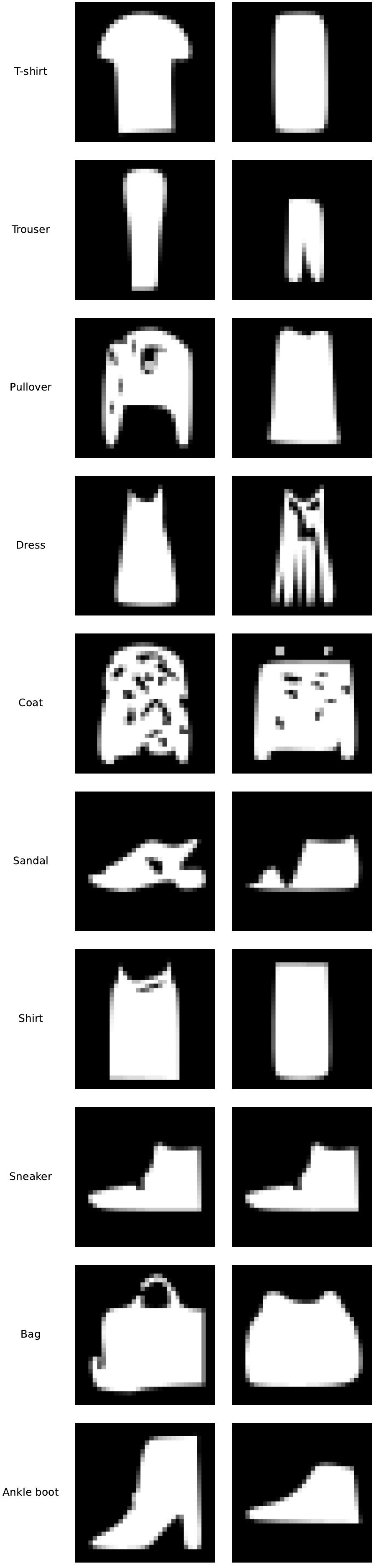}
    \caption{FashionMNIST}
    \end{subfigure}
    \begin{subfigure}[b]{0.19\textwidth}
    \centering
    \includegraphics[width=\linewidth]{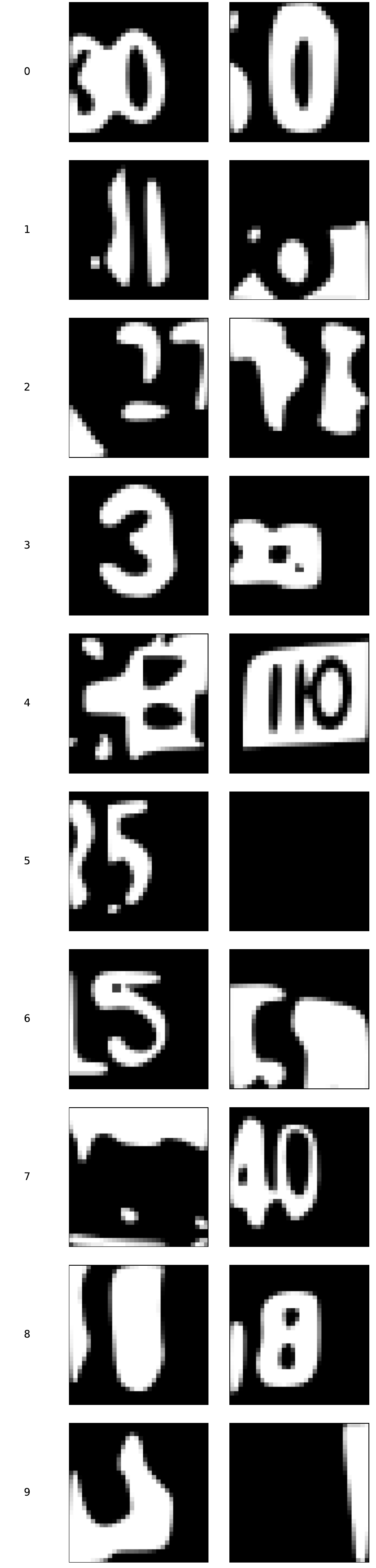}
    \caption{SVHN}
    \end{subfigure}
    \caption{DeepJanus original vs final candidate examples.}\label{fig:dj_row_comp}
\end{figure}
\end{document}